\documentclass{aa}
\usepackage[varg]{txfonts}
\usepackage{booktabs,graphicx,siunitx}
\DeclareSIUnit \parsec {pc}
\DeclareSIUnit \pc     {\parsec}
\DeclareSIUnit \kpc    {\kilo \parsec}
\DeclareSIUnit \msun   {\ensuremath{M_\sun}}
\DeclareSIUnit \year   {a}
\newcommand{\Al}{\element[][26]{Al}\xspace}
\usepackage{hyperref}

\begin{document}

\title{Kinematics of massive star ejecta in the Milky Way\\
  as traced by \textsuperscript{26}Al}

\author{
  Karsten Kretschmer   \inst{\ref{inst:apc}, \ref{inst:mpe}} \and
  Roland Diehl         \inst{\ref{inst:mpe}, \ref{inst:xcu}} \and
  Martin Krause        \inst{\ref{inst:mpe}, \ref{inst:xcu}} \and
  Andreas Burkert      \inst{\ref{inst:usm}, \ref{inst:xcu},
                             \ref{inst:mpe}} \and
  \\
  Katharina Fierlinger \inst{\ref{inst:xcu}, \ref{inst:usm}} \and
  Ortwin Gerhard       \inst{\ref{inst:mpe}} \and
  Jochen Greiner       \inst{\ref{inst:mpe}, \ref{inst:xcu}} \and
  Wei Wang             \inst{\ref{inst:nao}}
}

\institute{
  François Arago Centre, APC, Université Paris Diderot,
  CNRS/IN2P3, CEA/Irfu, Observatoire de Paris, Sorbonne Paris Cité,\\
  10 rue Alice Domon et Léonie Duquet, 75205 Paris Cedex 13, France
  \label{inst:apc}\\
  \email{kkretsch@apc.univ-paris7.fr}
  \and
  Max-Planck-Institut für extraterrestrische Physik,
  Giessenbachstrasse 1, 85741 Garching, Germany
  \label{inst:mpe}
  \and
  Excellence-Cluster ``Origin \& Structure of the Universe'',
  Boltzmannstrasse 2, 85748 Garching, Germany
  \label{inst:xcu}
  \and
  Universitätssternwarte der Ludwig-Maximilians-Universität,
  Scheinerstrasse 1, 81679 München, Germany
  \label{inst:usm}
  \and
  National Astronomical Observatories, Chinese Academy of Sciences,
  Beijing 100012, China
  \label{inst:nao}
}

\date{Received 29 August 2013 / accepted 19 September 2013}

\abstract
{Massive stars form in groups and their winds and supernova explosions
  create superbubbles up to \si{\kpc} in size. The fate of their
  ejecta is of vital importance for the dynamics of the interstellar
  medium, for chemical evolution models and the chemical enrichment of
  galactic halos and the intergalactic medium. However, ejecta
  kinematics and the characteristic scales in space and time are
  rather unexplored beyond \SI{\sim 10}{\kilo \year}.}
{Through measurement of radioactive \Al with its decay time constant
  of \SI{\sim e6}{years}, we aim to trace the kinematics of cumulative
  massive-star and supernova ejecta independent of the uncertain gas
  parameters over million-year time scales. Our goal is to identify
  the mixing time scale and the spatio-kinematics of such ejecta from
  the \si{\pc} to \si{\kpc} scale in our Milky Way.}
{We use the SPI spectrometer on INTEGRAL and its observations along
  the Galactic ridge to trace the detailed line shape systematics of
  the \SI{1808.63}{\keV} gamma-ray line from \Al decay. We determine
  line centroids and compare these to Doppler shift expectations from
  large-scale systematic rotation around the Galaxy's center, as
  observed in other Galactic objects.}
{We measure the radial velocities of gas traced by \Al, averaged over
  the line of sight, as a function of Galactic longitude. We find
  substantially higher velocities than expected from Galactic
  rotation, the average bulk velocity being $\SI{\sim 200}{\km\per\s}$
  larger than the Galactic-rotation prediction. The observed radial
  velocity spread implies a Doppler-broadening of the gamma-ray line
  that is consistent with our measurements of the overall line width.
  We can reproduce the observed characteristics with \Al sources
  located along the inner spiral arms, when we add a global blow-out
  preference into the forward direction away from arms into the
  inter-arm region, such as expected when massive stars are offset
  towards the spiral-arm leading edge. With the known connection of
  superbubbles to the gaseous halo, this implies angular-momentum
  transfer in the disk-halo system and consequently also radial gas
  flows. The structure of the interstellar gas above the disk affects
  how ionizing radiation may escape and ionize intergalactic gas.}
{}


\maketitle

%
\section{Introduction}
\label{sec:intro}

Massive stars are important agents of the evolution of gas and stellar
content in a galaxy, as they evolve rapidly within millions of years
(\si{\mega \year}) and are powerful sources of energy through their
ionizing radiation, their winds, and the final supernova explosions
\citep{zinnecker07:_towar_under_massiv_star_format}. Mostly formed in
groups \citep{lada03:_embed_clust_molec_cloud}, they create
superbubbles up to \si{\kpc} in size
\citep{jaskot11:_obser_const_super_x_energ_budget, weaver77:_inter},
and drive large scale outflows \citep{glasow13:_galac_lyman}. Ejecta
transfer their kinetic energy in a complex way to the structured
interstellar gas. The global picture of how the gas, metal and energy
output of these massive stars in the form of stellar winds and
supernova ejecta interacts with their surroundings is still unclear.
Our knowledge of the transport of gas and energy is derived from
measurements of observables of different types, each with biases and
imperfections: Dense molecular gas seen in CO
\citep{dame01:_milky_way_co}, atomic gas through \ion{H}{I}
\citep{kalberla06:_global_hi}, X-ray emission partly in interacting
shells and from hot cavity interiors
\citep{snowden97:_rosat_survey_diffus_x_ray_backg_maps}, free-free
emission from decomposition of radio emission
\citep{bennett96:_four_year_cobe_dmr}, and gamma-rays from decays of
unstable isotopes tracing nucleosynthesis ejecta
\citep{voss09:_using_ob}.  The initial release of matter and energy is
reflected in supernova remnants, which can be studied in a variety of
wavelength regimes over up to several \SI{10000}{years}; but
thereafter, radiative effects of the interstellar impacts from massive
stars fade away. Only long-lived radio-isotopes then provide a new and
different type of radiation, observable through a radioactive
afterglow in characteristic gamma-rays over millions of years from \Al
($\tau\sim\SI{e6}{\year}$) and \element[][60]{Fe}
($\tau\sim\SI{3.8e6}{\year}$). Here we report on observations of \Al
through its characteristic gamma-ray line at an energy of
\SI{1808.63}{\keV}, which has been measured with the SPI telescope
\citep{vedrenne03:_spi} on the INTEGRAL satellite
\citep{winkler03:_integ}.

Earlier analysis of such observations had provided hints of systematic
Doppler shifts of the \Al gamma-ray line with Galactic longitude,
which were consistent with large-scale Galactic rotation
\citep{diehl06:_radioac_al_galax}.  This showed that \Al is sampled
throughout the Galaxy with such gamma-ray line observations, as
gamma-rays penetrate even molecular clouds which may be assumed to
surround some of the youngest source regions. Comparing \Al emission
with the spatial distribution of candidate sources, it has been
confirmed that groups of massive stars are the most-plausible origins
\citep{prantzos96:_26al_galaxy}. Among several tracers of \Al sources,
diffuse emission from ionized gas through free-free emission has been
found most promising \citep{knoedlseder99:_comptel_26al}, although
diffuse dust emission or cosmic-ray interactions with interstellar gas
as seen in continuum gamma-ray emission also provide a good
correlation to \Al emission. Such studies of correlations between the
angular distribution of different observables are limited by the
spatial resolution achieved in the \SI{1808.63}{\keV} line, which does
not exceed 3~degrees in any existing measurement. For very nearby
sources such as the Orion OB1 subgroups, COMPTEL observations had
suggested that \Al emission may be offset from its sources and rather
arise from extended emission in a cavity blown by earlier activity of
the massive-star association
\citep{diehl03:_nucleo_massiv_stars_orion_region}. Such superbubbles
may have an important role in transport of energy and ejecta from
their sources back into interstellar-medium phases which may form
stars again. A hint towards this also may be derived from indications
that the \Al emission scale height perpendicular to the Galactic plane
of \SI{\sim 130}{\pc} appears to fall on the high side of scale
heights which characterize the \Al sources (e.g.\ molecular gas
measured in CO has \SI{\sim 50}{\pc} scale height)
\citep{wang09:_spect_galac_al}.

With accumulating exposure, we now extend our study of \Al throughout
the Galaxy, to better trace and compare the kinematics of \Al in the
inner \SIrange{4}{5}{\kilo\parsec} of our Galaxy
(Figs~\ref{fig:spectra-ref}, \ref{fig:lvd}).  We present the methods
we used to obtain these measurements and discuss the sources of
uncertainty involved (Sect.~\ref{sec:spi-analysis}). We then discuss
our longitude-velocity measurements in relation to previous
measurements of longitude-velocity dependence obtained using other
observables, such as CO (Sect.~\ref{sec:results}).  We find that our
kinematic results support the notion that superbubbles are the
structures which are most important in transport of energy and ejecta
on the longer (\si{\mega \year}) time scales, which are characteristic
for recycling of matter and energy.  Towards further interpretations,
we then present a first-order spatio-kinematic model capable of
explaining the differences between those and our \Al data
(Sect.~\ref{sec:discuss}).

%
\section{Data and their Analysis}
\label{sec:spi-analysis}

%
\subsection{Mission and Data}
\label{sec:spi-mission}

The INTEGRAL space observatory \citep{winkler03:_integ} carries the
gamma-ray spectrometer instrument SPI as one of its two main
instruments \citep{vedrenne03:_spi,roques03:_spi_perf}. The SPI
spectrometer features a camera consisting of 19 high-resolution Ge
detectors, which measures celestial gamma-rays through coded-mask
shadowgrams, above a large instrumental background.  SPI data consist
of energy-binned spectra for each of the 19 Ge detectors of the SPI
telescope camera \citep{vedrenne03:_spi}, taken in typically
\SI{30}{\minute} exposures of a sky region.  For our analysis, we used
exposures across the plane of the Galaxy accumulated over more than
\SI{9}{years} of the INTEGRAL mission \citep{winkler03:_integ,
  winkler11:_integ}.

\begin{figure}
  \centering
  \includegraphics[width=\linewidth]{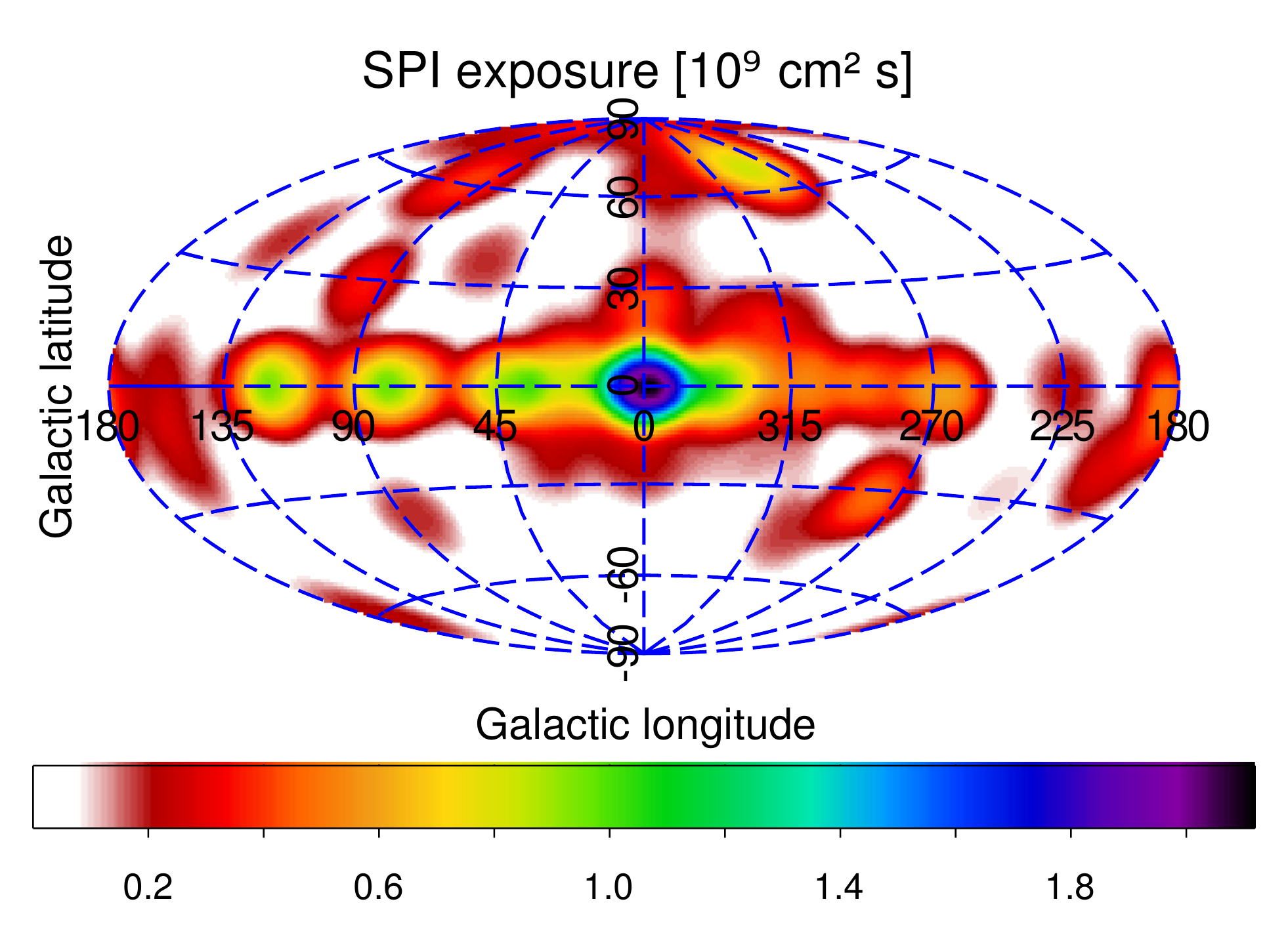}
  \caption{Exposure map of the sky with the SPI telescope on INTEGRAL,
    for the data used in this analysis (Feb 2003 to Feb 2012).}
  \label{fig:spimodfit-sa-exposuremap}
\end{figure}

\begin{figure}
  \centering
  \includegraphics[width=\linewidth]{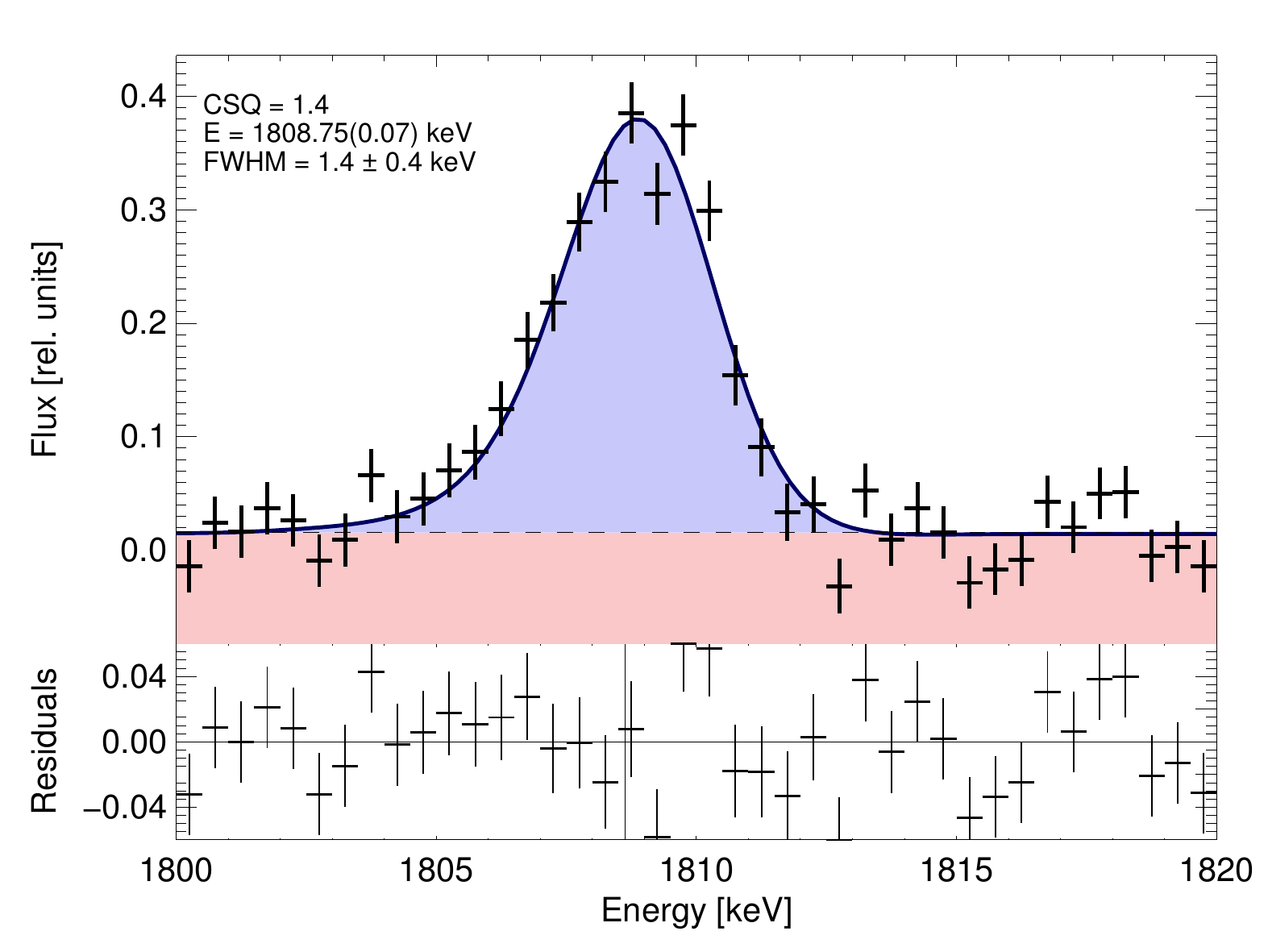}
  \caption{Spectrum around the gamma-ray line from \Al, as obtained
    for the entire inner Galactic plane ($\text{ROI} = \SI{128 x
      12}{\degree}$ at $l = \ang{0}$) }
  \label{fig:spec-dl128db12}
\end{figure}

%
\subsection{Generating spectra of celestial emission}
\label{sec:spi-gamma}

Generally, in our spectroscopy analysis we fit the intensity scaling
factor of a model of the \Al sky intensity distribution plus a scaled
model of the instrumental background to the set of spectra accumulated
during multi-year observations from our 19 Ge detectors and instrument
pointings.  Our 9-year observations set includes \num{65302}
instrument pointings that add up to the exposure shown in
Fig.~\ref{fig:spimodfit-sa-exposuremap}.  Data $d_k$ are modelled as a
linear combination of the sky model components $M_{ij}$, to which the
instrument response matrix $R_{jk}$ is applied, and the background
components $B_{jk}$:
\begin{equation}
  d_k = \sum_j R_{jk} \sum_{i = 1}^{N_\mathrm{I}}
  \theta_i M_{ij} + \sum_{i = N_\mathrm{I} + 1}^{N_\mathrm{I} +
    N_\mathrm{B}} \theta_i B_{jk}\label{eq:model-fit}
\end{equation}
i.e.\ the comparison is performed in data space, which consists of the
counts per energy bin measured in each of SPI's detectors for each
single exposure of the complete observation.

By way of the $M_{ij}$ terms, we make use of prior knowledge in the
form of a sky intensity distribution such as e.g.\ the measured \Al
intensity, or a plausible model such as an exponential disk. We use
the \SI{1.8}{\MeV} sky map from the COMPTEL gamma-ray telescope
\citep{schoenfelder93:_comptel} on the NASA CGRO mission (1991-2000)
\citep{plueschke01:_compt} as derived through maximum-entropy
deconvolution \citep{strong95:_maxim_entrop_imagin_compt_data}. Our
analysis also includes a model for the behaviour of the instrumental
background, which is derived from separate analysis of the continuum
intensity in energy bands adjacent to the \SI{1808.63}{\keV} line and
instrumental background tracers in data of the entire INTEGRAL
mission. The result of such background study on independent data is a
prediction of counts per detector, energy bin, and spacecraft
pointing, which is adjusted to the data together with the predicted
contribution from the sky (i.e.\ the sky intensity model as folded
into data space using the instrument imaging response function).  We
then repeat this for \SI{0.5}{\keV} wide energy intervals to obtain
the sky intensity spectra for such an adopted sky distribution
model. In Fig.~\ref{fig:spec-dl128db12} the spectrum is shown for the
entire inner Galaxy, while in Fig.~\ref{fig:spectra-ref}, different
spectra are shown for spatially separated regions of the sky.  The
background and sky models and fitting method used in this step are
identical to previous work \citep{wang09:_spect_galac_al} and
summarised briefly below.

\begin{figure}
  \centering
  \includegraphics[width=\linewidth]{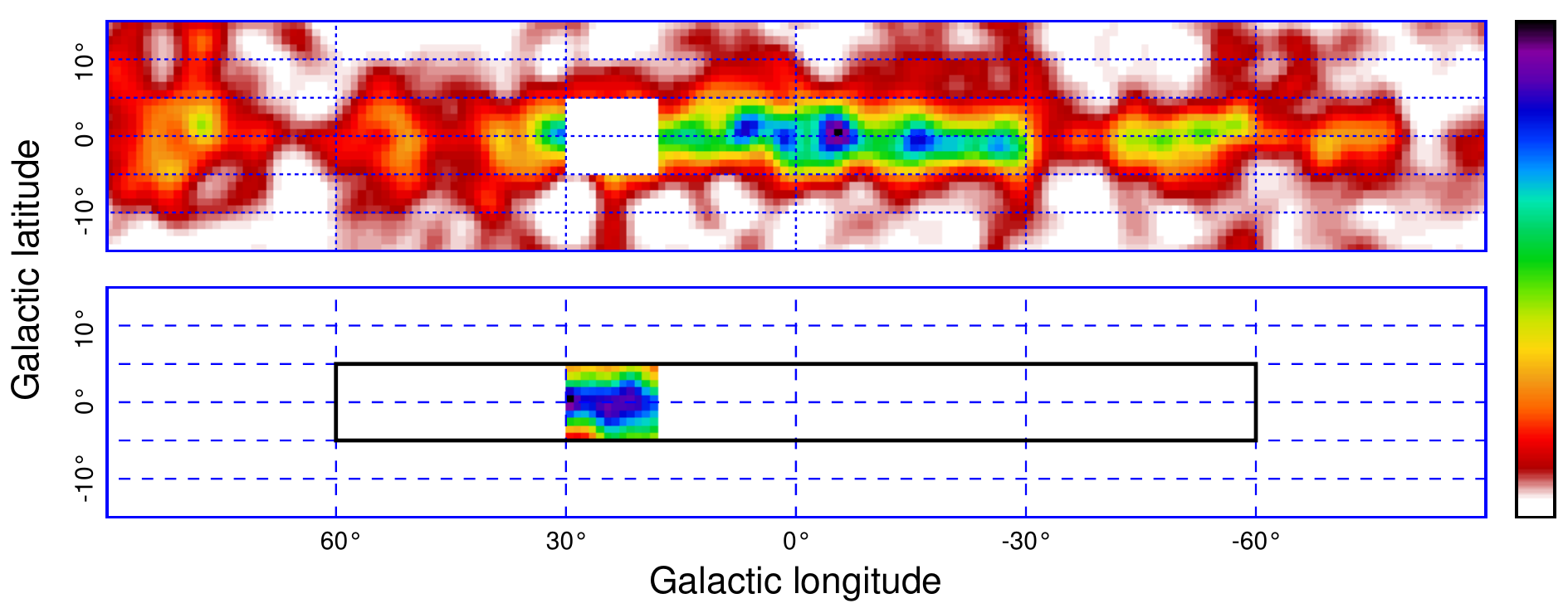}
  \caption{Two sky model components: The region of interest (``ROI'')
    \SI{12 x 10}{\degree} centred around $l = \ang{24}$ (bottom) and
    its complement (top), taken as subsets of the COMPTEL
    \SI{1.8}{\MeV} sky map. The rectangular outline shows the region
    covered by our scan along the Galactic plane.}
  \label{fig:spimodfit-sa-maps}
\end{figure}

In order to improve in line Doppler shift sensitivity compared to
previous SPI results \citep{diehl06:_radioac_al_galax}, we implemented
a new approach for scanning the Galactic plane, employing sky models
which are split into two independent components. The sky model we use
(e.g.\ the \Al observed with COMPTEL, \citealp{plueschke01:_compt}),
is divided into two complementary parts: the inside of the spherical
rectangle $l \in [l_0 - \Delta l/2,\ l_0 + \Delta l/2]$, $b \in [b_0 -
\Delta b/2,\ b_0 + \Delta b/2]$ defines our region of interest
(``ROI''), and its complement with respect to the full-sky map
constitutes the second component of the model
(Fig.~\ref{fig:spimodfit-sa-maps}).  A full-sky model is required,
because SPI observation data include events from within the entire
telescope field of view of \ang{\sim 30} extent, although the
intrinsic spatial resolution of SPI has been determined as \ang{2.7}.
The sky model thus represents the spatial detail of the fitted
intensity within a longitude/latitude bin (ROI); the intensity is
fitted to SPI data for the entire ROI; spatial details within ROI bins
have little impact on the spectral-line results due to the low total
\Al signal per ROI. This was confirmed, using different plausible \Al
sky maps with different structural detail on the scale below
$\sim$~few degrees.

\begin{figure}
  \centering
  \includegraphics[width=\linewidth]{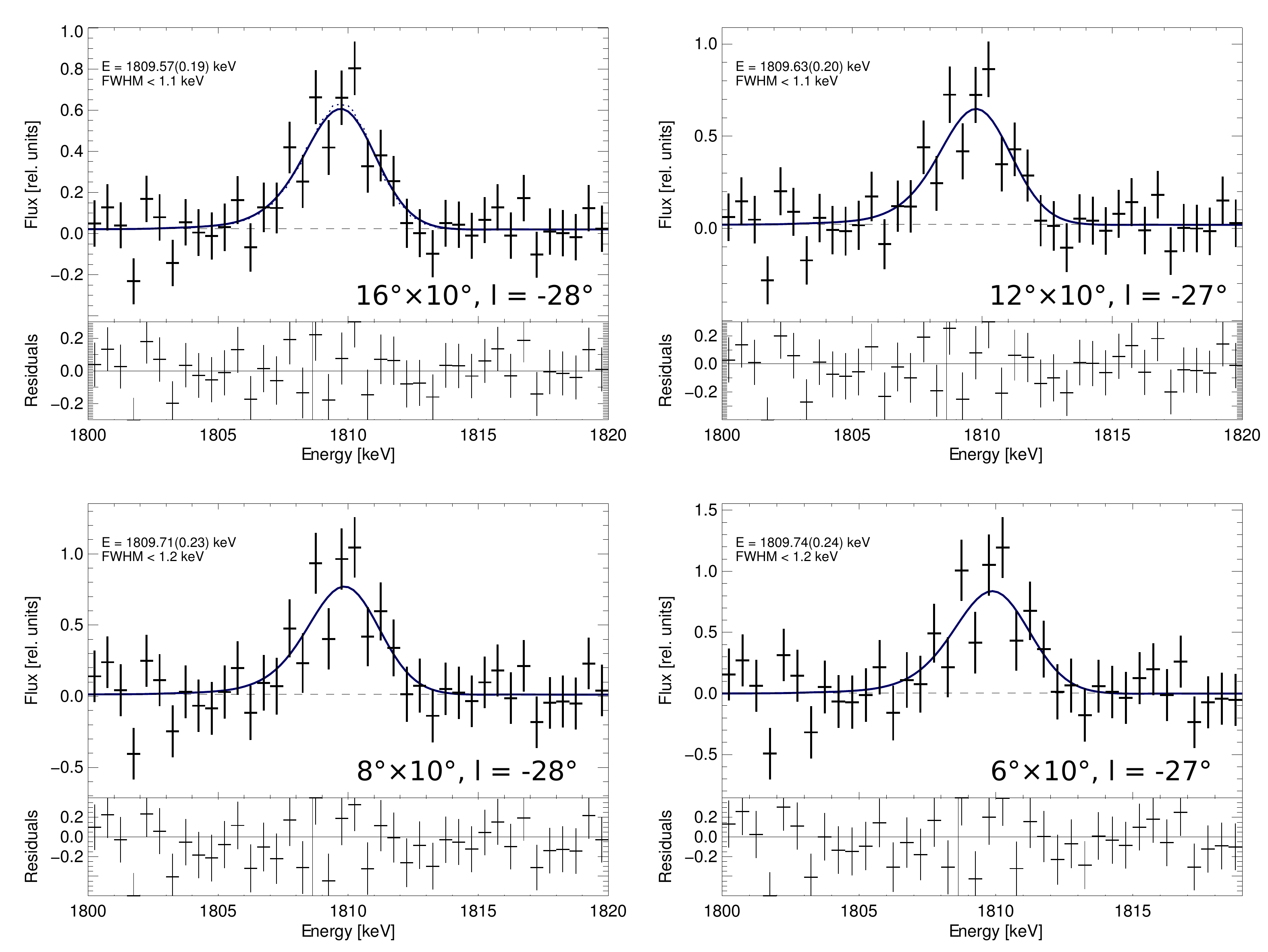}
  \caption{Spectra of the region around $l \sim
    {}$\SIrange{-27}{-28}{\degree} for ROI latitude extents of
    \SIlist{16;12;8;6}{\degree}}
  \label{fig:spimodfit-sa-4spec}
\end{figure}

The intensities of these two components, together with a model of the
instrumental background, are then fitted to the SPI data. Our
background model reproduces the time variability of the background at
short timescales (\SI{< 3}{days}) with the rate of events saturating
the germanium detectors, which has been found to be a sensitive
measure of the instantaneous charged particle environment of the
instrument. The long term background variation (\SI{> 3}{days}) is
extracted from the continuum intensity in energy intervals adjacent to
the \SI{1808.63}{\keV} line.

Thus we obtain spectra in the energy range \SIrange{1800}{1820}{\keV}
around the \Al line for the two complementary sky model components. We
repeat this process, varying $l_0$ to scan the ROI along the Galactic
plane, and obtain measurements of the \Al line signal as a function of
Galactic longitude.  Fig~\ref{fig:spimodfit-sa-4spec} shows sample
results for a particular such region-of-interest in the fourth
quadrant of the Galaxy towards longitude $l \sim
{}$\SIrange{-27}{-28}{\degree} (the centre longitude is not identical
because of the \SIlist{4;3;2;3}{\degree} rasters -- one quarter or one
half of the longitude extent -- used for the different respective
longitude extents).

The latitude range $\Delta b = \ang{\pm 5}$ in our analysis was chosen
to cover the full expected scale height for both ejecta as well as gas
streaming away from the plane of the Galaxy towards the halo even for
nearby segments of the Galaxy. This is equivalent to \SI{\pm 270}{\pc}
at \SI{3}{\kpc} distance; the CO disk scale height is \SI{\sim
  50}{\pc} \citep{dame01:_milky_way_co}, a previous \Al scale height
estimate \citep{wang09:_spect_galac_al} finds a range \SIrange{\sim
  60}{\sim 250}{\pc}. But foreground emission, which would
predominantly be showing up at intermediate or higher latitudes, may
lead to possible biases: The ROI, which corresponds to a pyramid in
3D-space, covers different distances from the Galactic plane,
depending on the distance to the emitting region. Nearby sources,
taking up a large solid angle on the sky across the plane would thus
be sampled only partially, depending on the ROI latitude extent. The
influence of the choice of ROI longitude and latitude extent on the
model fit results is discussed in greater detail in
Appendix~\ref{sec:lvd-analysis_errors}.

\begin{figure}
  \centering
  \includegraphics[width=\linewidth,viewport=28 6 328 212]{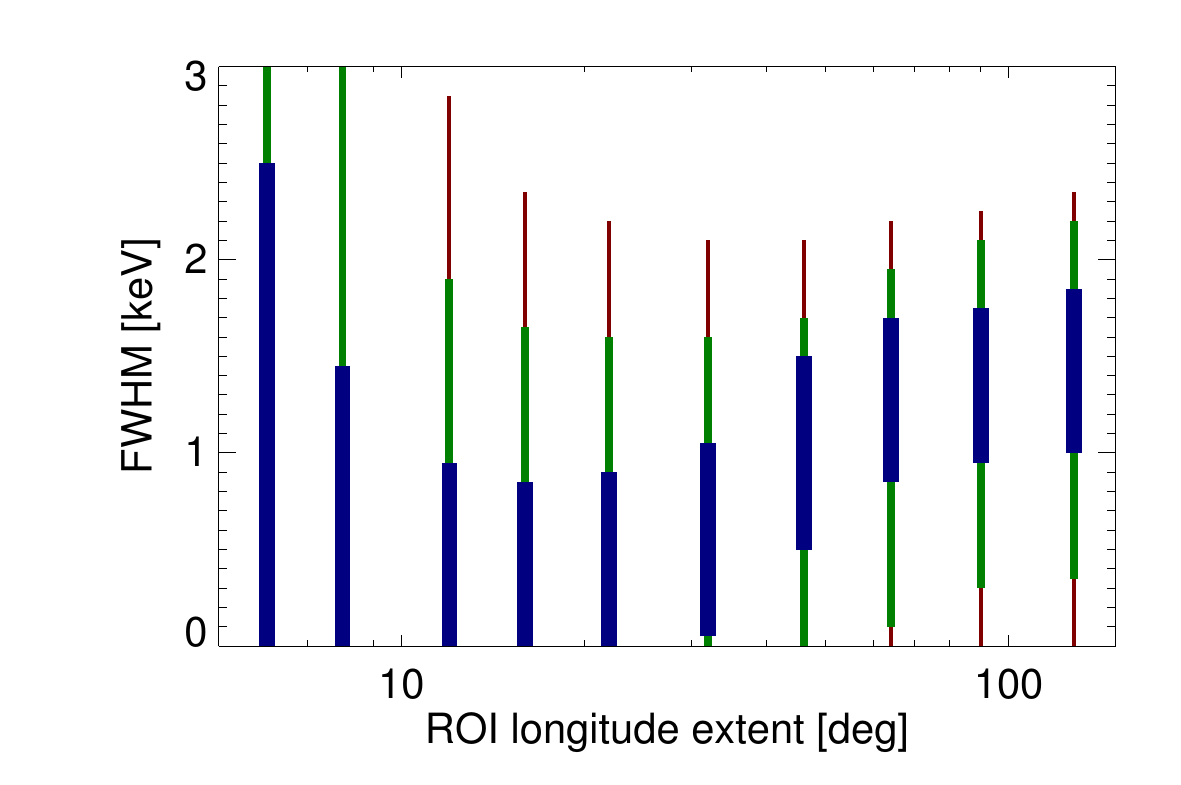}
  \caption{Limits on the width of the \Al line emission from ROIs
    around the Galactic centre with an extent in latitude of \ang{+-6}
    and longitude extents ranging from \SIrange{6}{128}{\degree}. The
    error bars of decreasing thickness show the (Bayesian) highest
    posterior density credible regions corresponding to
    \SIlist[round-mode = figures, round-precision =
    3]{68.268950;95.449974;99.730020}{\percent}, corresponding to
    \SIlist{1;2;3}{\ensuremath{\sigma}}.}
  \label{fig:igl-hpdcr}
\end{figure}

The spread of radial velocities over the longitude range covered
implies an overall broadening of the line emission when considering
the integrated emission coming from a large ROI on the sky. When we
vary the extent of the extent of the ROI and with it the range of
radial velocities being integrated over, and compare this to the
measured line width, we can check the broadening effect of different
samples of the sky and its varying kinematic properties.  This is
shown in Fig.~\ref{fig:igl-hpdcr}, where the error bars show the
variation of the measured line width's confidence intervals with
longitude extent. For small regions on the sky, only upper limits on
the \Al line width can be obtained. The upper limits become smaller
with increasing region size as more signal is covered by the
region-of-interest before increasing again for $\Delta l \la \ang{30}$
as the covered radial velocity range increases. For the large region
along the entire inner Galaxy (\SI{128 x 12}{\degree},
Fig.~\ref{fig:spec-dl128db12}), we derive \SI{1.4 \pm 0.4}{\keV}
(FWHM) additional broadening, or \SI{230 \pm 70}{\km\per\s}. This is
consistent with the spread of radial velocities we measure spatially
resolved along the inner Galaxy (Sect.~\ref{sec:lvd-analysis}). The
root-mean-square of our radial velocity measurements
(Fig.~\ref{fig:lvd}), weighted with the corresponding intensity
measurements (Fig.~\ref{fig:inten-db}) is \SI{\approx
  200}{\km\per\s}. We thus conclude that observed line broadenings are
consistent with systematic variation of line position along the plane
of the Galaxy, attributed to large-scale rotation of gas within the
Galaxy.

%
\subsection{Characterizing celestial emission lines}
\label{sec:spi-gammashape}

In a second step, we fit these spectra of sky intensity values
obtained per energy bin and per component by a model description based
on the instrument's spectral response. This yields the \Al line
parameters of total intensity, Doppler shift with respect to
laboratory energy, and intrinsic width of the celestial \Al emission,
for the respective component of the sky.  The spectral model we use in
our line fitting consists of a linear continuum and a line at the
position $a_3$, where the line is the convolution of the instrument
spectral response $R$ and a Gaussian $G$ with the width $a_4$ ($E_0$
is the midpoint of the energy interval):
\begin{equation}
  I(E) = a_0 + a_1 (E - E_0) + a_2 (R * G_{a_4})(E -
  a_3)\label{eq:spec-model}
\end{equation}
Such detailed modelling of the instrument response is required, as the
impact of cosmic radiation onto SPI detectors gradually deteriorates
the charge collection properties of detectors, and leads to a degraded
spectral response.

\begin{figure}
  \centering
  \includegraphics[width=\linewidth]{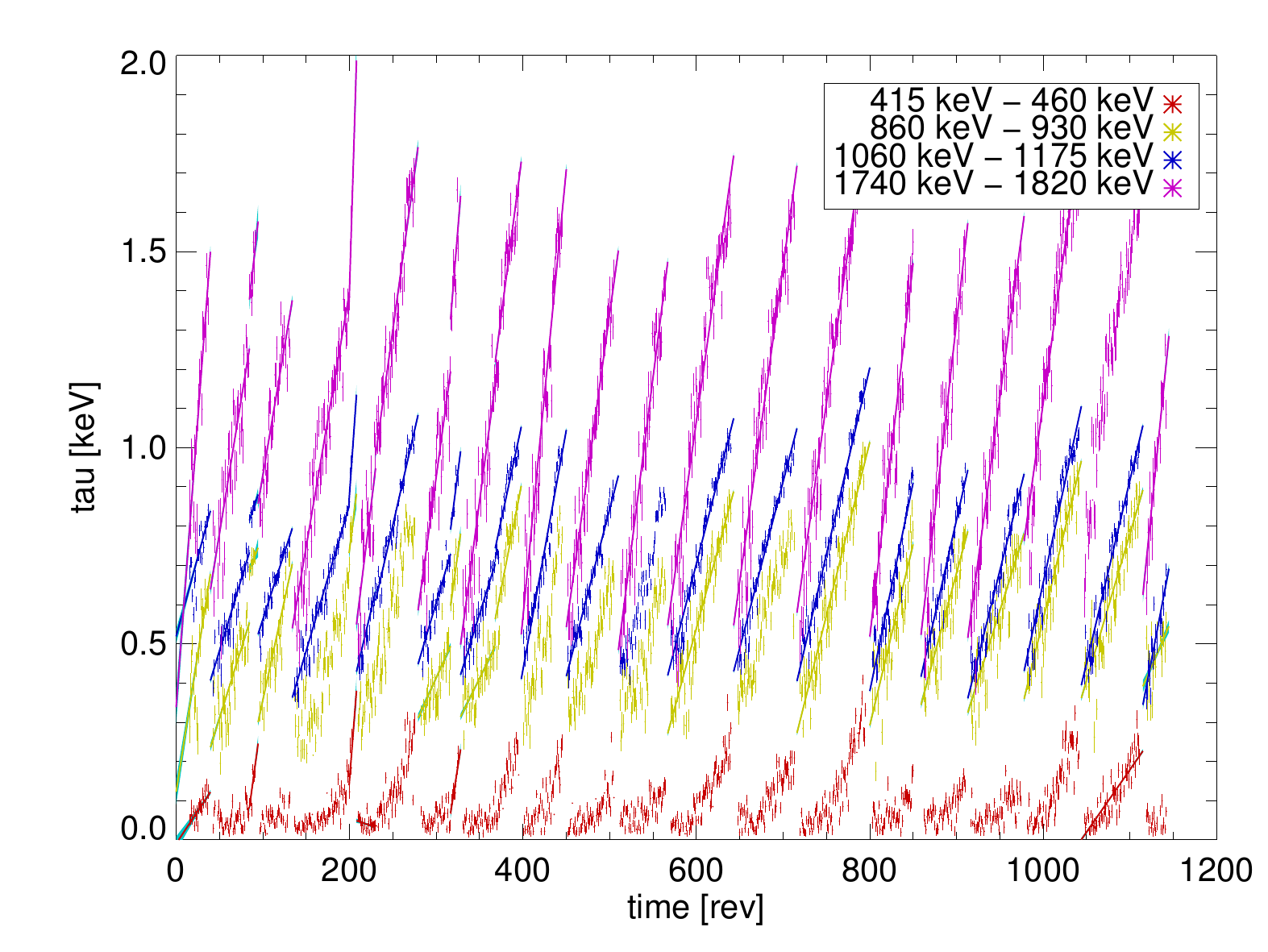}
  \caption{Degradation of the spectral response over mission
    time. With degradation, a one-sided tail at the low-energy side
    develops. The plot shows its extent versus time (one revolution =
    \SI{3}{days}).}
  \label{fig:cal-tau}
\end{figure}

The degradation of Ge detectors from cosmic-ray irradiation and its
restoration in annealings results in a time variable width and
asymmetry of the spectral response. This variation of the spectral
response dominates over all other spectral changes, and is found
consistent across the SPI energy range. Figure~\ref{fig:cal-tau} shows
how the degradation -- as measured by the width of a one-sided
exponential tail on the low-energy side of the line response -- varies
for lines in four different energy regimes.  The degradation increases
in an approximately linear fashion with time and is reduced
periodically by the annealing operations. These heat the detectors for
a period of \num{\sim 2} weeks to \SI{\sim 100}{\celsius}, restoring
the original high spectral resolution. The annealing cycle of one
roughly every 6 months (as needed) lead to a sawtooth-like variation
of the spectral response during our data taking.  Clearly, the
absolute magnitude of degradation increases with energy, yet changes
occur consistently for all instrumental lines, and are in the range of
tenths of \si{\keV}.  Spectroscopic analysis at high precision needs
to account for these effects.

We estimate the spectral-model parameters (continuum intensity and
slope; line intensity, Doppler shift, and width) using the
Metropolis-Hastings MCMC algorithm
\citep{neal93:_probab_markov_monte_carlo}, which samples the parameter
space statistically to generate detailed probability distributions for
the model parameters. This approach allows us to determine accurate
limits on the width of the emission line even though the differences
between the measured line profiles and the instrumental line response
are small. Figure~\ref{fig:spectra-ref} compares the observed Doppler
shifts of the \Al line to the spectral appearance of a nearby
instrumental line for exactly the same data, i.e.\ all selected SPI
pointings around the respective longitude ROI.  It is evident that the
energy calibration is stable, and the relative changes of the response
are small (see Sect.~\ref{sec:lvd-analysis} below and
Fig.~\ref{fig:cal-tau}). The MCMC analysis also determines the Bayes
factor \citep{gelman03:_bayes}, i.e.\ the ratio of the marginal
likelihoods of the model including a line and the alternative model
that consists of continuum only. If the Bayes factor is larger than
one, the model with a line is more probable than the one without,
although the line detection may still be insignificant in terms of
statistical acceptance criteria. Because of the large numbers
involved, it is more convenient to state the Bayes factor on the
logarithmic decibel (\si{\dB}) scale. The longitude-velocity graph
(Fig.~\ref{fig:lvd}) shows the result of this process for $\Delta l =
\ang{12}$, $\Delta b = \ang{10}$. For our \SI{12 x 10}{\degree} ROI,
the Bayes factor reaches a maximum of \SI{135}{\dB}, which corresponds
to a detection significance of \SI{7.6}{$\sigma$}. For the sky
distribution model from the \Al emission as derived with COMPTEL, and
evaluated over the region $|l| < \ang{60}$, $|b| < \ang{30}$, the
Bayes factor and detection significance values are \SI{2100}{\dB} and
\SI{31}{$\sigma$}, respectively. Therefore, for sky region sizes as we
use here in our ROIs with \SI{\sim 120}{square degrees} or less, the
line widths are not tightly constrained, and we give upper limits
only; these are: \SI{< 1.9}{\keV} (i.e.\ \SI{< 315}{\km\per\s};
\SI{95.4}{\percent}, \SI{2}{$\sigma$}), or \SI{< 2.75}{\keV} (\SI{<
  455}{\km\per\s}; \SI{99.7}{\percent}, \SI{3}{$\sigma$}). Only
integrating the signal over a larger solid angle allows a more precise
line width measurement (Fig.~\ref{fig:spec-dl128db12} and
Sect.~\ref{sec:lvd-analysis}).

The basic instrument spectral response is obtained from the
instrumental background line at \SI{1764.494}{\keV}, which is due to
the \element[][214]{Bi} $\beta$-decay inside SPI's anticoincidence
shield. This response is shown in Fig.~\ref{fig:spectra-ref}
(right-hand set of spectra). Its shape is nearly Gaussian, it exhibits
however an excess towards the low-energy side that is due to partial
charge collection and which varies with the degradation state of the
detectors between annealings \citep{roques03:_spi_perf}, as described
above.

For a given ROI position, we determine the spectral response from all
exposures where SPI was pointed at a location within a field-of-view
radius (\ang{16}) around the ROI. We add the spectral responses of the
active SPI detectors, we also adjust the absolute energy scale to
compensate for the radial velocity due to the Earth's orbital motion,
further adjust the energy scale relative to the line centroid to
extrapolate the widening of the instrument energy response with
increasing photon energy. We sum these per-exposure spectra, subtract
the linear continuum and normalize the result to obtain the SPI energy
response for this ROI.

We also tested a different spectral response determination, which uses
a parametrized analytic function (the convolution of a Gaussian with
an exponential defined on $x \le 0$) to describe the line shape, and
extracts the line shape parameters from a set of eleven strong
instrumental background lines with high time resolution (\SI{3}{days})
over the multi-year observation.  The impact of the choice of the
energy response model on the measured longitude-velocity dependence is
negligible, as shown in Fig.~\ref{fig:lvd-resp}: The RMS difference
for the longitude range $\ang{-36} \le l \le \ang{36}$ is
\SI{0.15}{\keV} (\SI{25}{\km\per\s}), i.e.\ small by comparison.
Appendix~\ref{sec:lvd-analysis_errors} discusses the systematic
uncertainties due to the instrumental response modelling in greater
detail.

%
\section{Results}
\label{sec:results}

%
\subsection{Longitude-velocity diagrams}
\label{sec:lvd-analysis}

\begin{figure}
  \centering
  \includegraphics[angle=90,width=\linewidth]{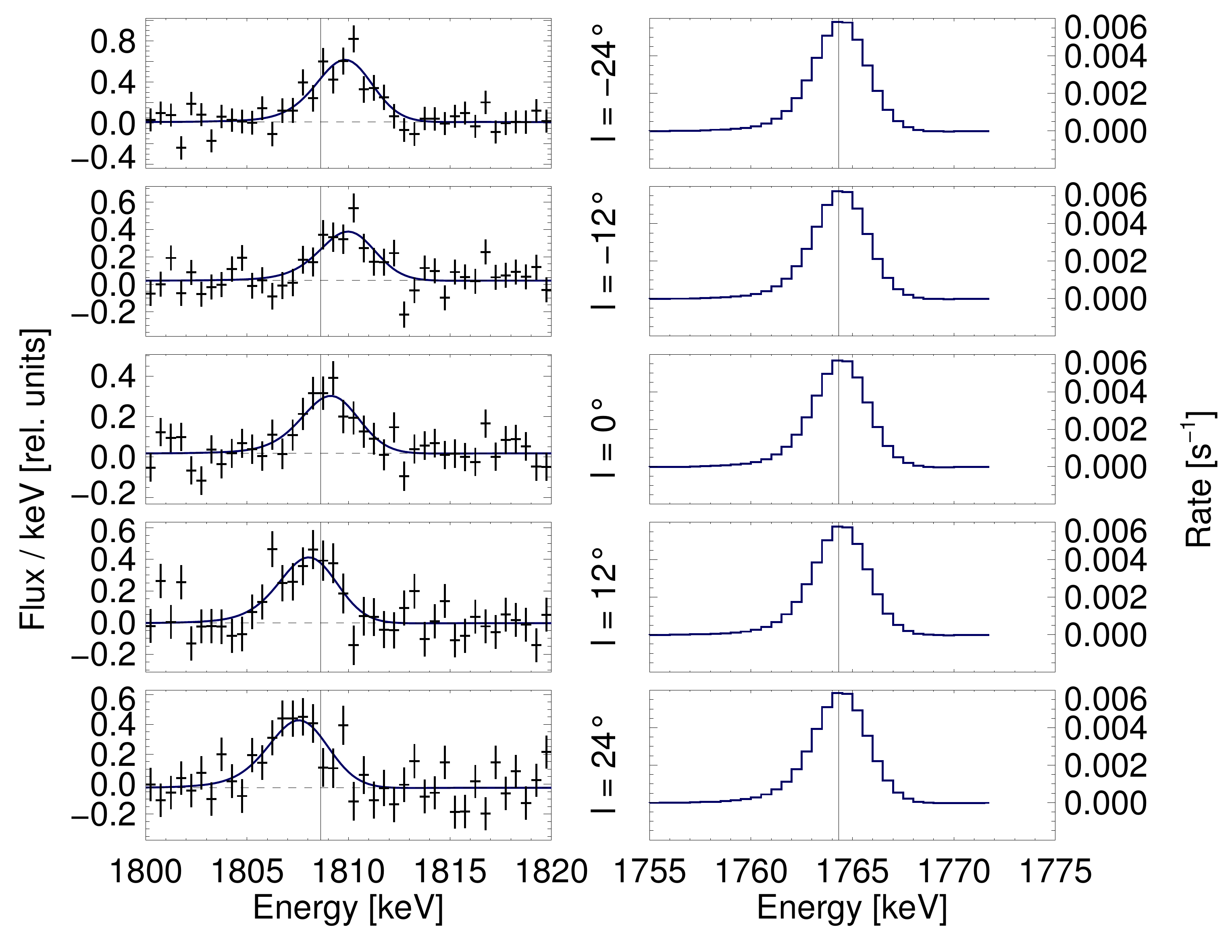}
  \caption{Trace of systematic Doppler shifts of the \Al line
    along the inner Galaxy. }
  \label{fig:spectra-ref}
\end{figure}

With the parametrization of the sky along the plane of the Galaxy in
bins of different Galactic longitude (at fixed latitude bin width), we
obtain spectra near the \Al line along the plane of the Galaxy
(Fig.~\ref{fig:spectra-ref}), as compared to a nearby instrumental
background line.  The line from decay of \Al (laboratory energy
\SI{1808.63}{\keV}) is seen for different regions along the plane of
the Galaxy (left), at galactic longitude, $l$,
\SIrange{-24}{24}{\degree}, for a latitude range $\Delta b = \pm
\ang{5}$ centred at $b = \ang{0}$. The shift of the line centroid with
galactic longitude is apparent, in particular when compared to an
instrumental line nearby (right; line at \SI{1764.494}{\keV}, from
activation of \element{Bi} in SPI's anticoincidence detector made of
BGO scintillator; background-line data are selected from the same
observations which contribute to the longitude bin shown on the
left). The instrumental line at \SI{1764.494}{\keV} demonstrates a
stable energy calibration for all observations, and in particular
absence of a bias with Galactic longitude. The line shape of the
instrumental line was used to represent the instrumental spectral
response, and used to fit the data in the left-side graphs, thus
determining the line position accurately even for weak signals.  The
systematic shift of the \Al line with Galactic longitude can clearly
be seen in this subset of spectra, which are selected from
non-overlapping, independent sky regions.

\begin{figure}
  \centering
  \includegraphics[width=\linewidth]{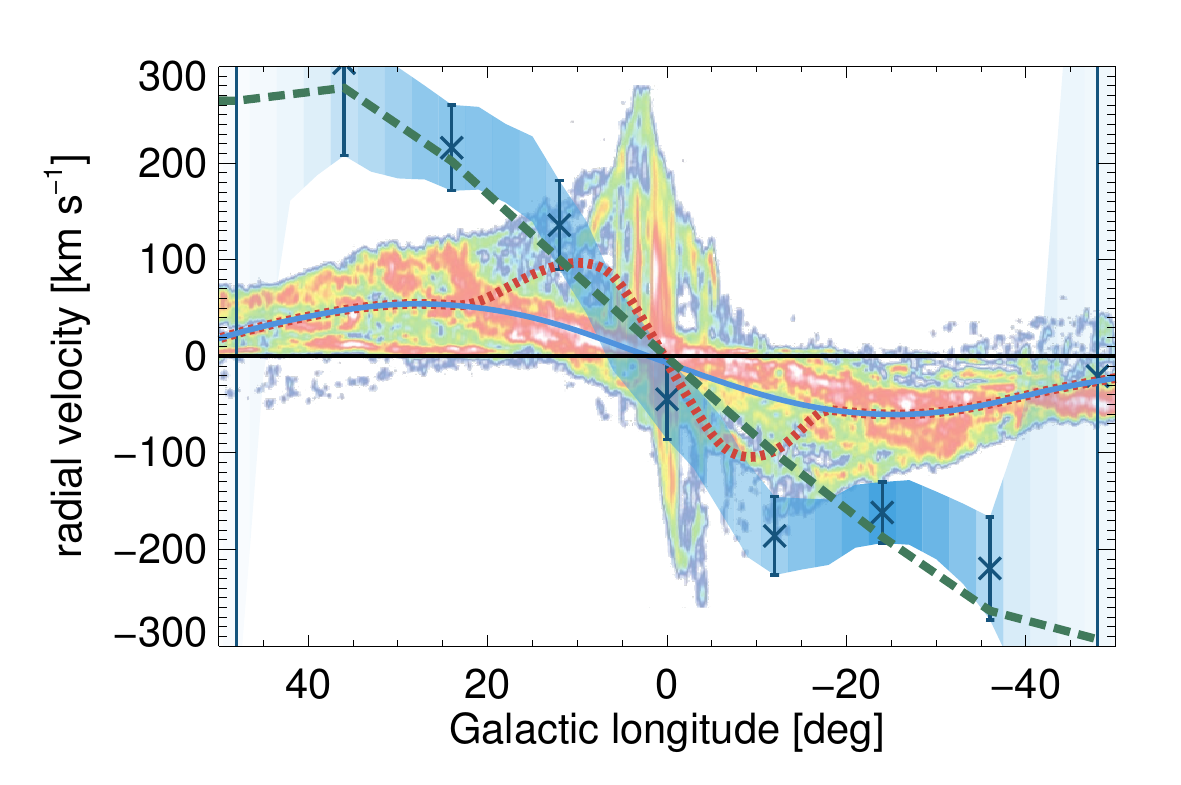}
  \caption{Longitude-velocity diagram comparing $\gamma$-ray-measured
    velocities (crosses, including error bars) with other objects in our Galaxy. \Al line-centroid
    energies were fitted to determine velocities in longitude bins of
    \SI{12}{\degree} and latitude ranges \ang{\pm 5}. For comparison,
    different models are shown (blue solid, red dotted, and green dashed lines), as
    well as the velocity information from molecular gas as seen in CO
    (details see Sect.~\ref{sec:discuss}). }
  \label{fig:lvd}
\end{figure}

We convert the measured offsets in the centroids of the \Al line from
the expected \Al decay at the laboratory value into the corresponding
bulk Doppler velocity. This allows us to construct a
longitude-velocity diagram from \Al measurements. Figure~\ref{fig:lvd}
shows the result, as derived from the spectra shown in
Fig.~\ref{fig:spectra-ref}.  This longitude-velocity result is
assembled from the above analysis, which we repeated for the
regions-of-interest (ROI) defined by rectangular bins in longitude
(widths \SIlist{6;8;12;16}{\degree}) and latitude (heights
\SIlist{4;6;8;10}{\degree}, Fig.~\ref{fig:lvd-dl}). Increasing the ROI
size trades spatial resolution against energy resolution. Our choice
of a \SI{12 x 10}{\degree} ROI offers balanced statistical and
systematic energy uncertainties (see
Appendix~\ref{sec:lvd-analysis_errors}).  By moving the ROI in
Galactic longitude, we can trace velocities along the plane of the
Galaxy. In the figure, we show data points (crosses with error bars
showing one standard deviation) spaced \ang{12} apart, i.e.\ offset by
integer multiples of the ROI width and therefore measuring
non-overlapping ROIs. We also show (in blue shading) measurements
obtained by a closer spaced \ang{3} longitude sampling which implies
that neighbouring ROIs overlap by \num{3/4}. This oversampling leads
to a stronger correlation between neighbouring measurements, but it
shows more information than the data points. The blue shaded areas
show a colour saturation proportional to the Bayes factor of the
spectral model compared to continuum-only, i.e.\ they show the
significance of the signal for each ROI position.  The systematic
blueshift in the 4\textsuperscript{th} and redshift in the
1\textsuperscript{st} Galactic quadrant are expected from large-scale
Galactic rotation.

%
\section{Discussion}
\label{sec:discuss}

Extraction of the spatio-kinematic characteristics of interstellar gas
in the inner Galaxy remains a challenge, due to distance ambiguities,
observational biases, and the model-dependence of velocity and
distance derivations -- in addition to the intrinsic differences in
resolution of different observables.  Our derived line centroids along
the Galactic ridge at $l \in [\ang{-50},\ang{50}]$
(Sect.~\ref{sec:lvd-analysis}) represent the average radial
velocities, subject to distance-dependent weighting, of \Al in volume
slices covering the whole inner Galactic plane.  As shown in
Fig.~\ref{fig:lvd}, we clearly find observed line-of-sight velocities
(relative to the local standard of rest) between approximately
\SI{200}{\km\per\s} (redshifted) and \SI{-200}{\km\per\s}
(blueshifted). The excess velocities beyond those globally expected
from Galactic rotation are \SI{100}{\km\per\s} and higher at the
velocity maxima, which are near longitudes $\pm\ang{30}$.

We believe the most likely explanation for our findings to be the
preferential expansion of superbubbles towards the leading edges of
spiral arms. This implies a net asymmetry of the million-year-scale
bubble expansions that results in a blow-out of massive star ejecta
into the low-density region ahead of and outward from the spiral arms.
For the kinematic description of \Al we must adopt a model for the
large-scale spatial distribution and kinematic behaviour of \Al as it
decays.  We now discuss different large-scale kinematic models, which
should explain the signature with longitude of \Al data points, and
finally support our suggested interpretation.

%
\subsection{Spatio-kinematic modelling of the \textsuperscript{26}Al
  longitude-velocity signature}
\label{sec:sk-model}

%
\subsubsection{Galaxy models with different large-scale rotation
  components}

In Fig.~\ref{fig:lvd}, we show in the continuous blue line what the
signature from \Al would be, if we assume that the density
distribution of the \Al in the disk and spiral arms of the Galaxy is
proportional to large-scale distribution of the free electron density
\citep{cordes02:_ne2001}, and that the gas is on circular orbits with
the velocity given by a Galactic rotation measurement compilation
\citep{sofue09:_unified_rotat_curve_galax}.  This line shows what our
gamma-ray telescope would have measured in longitude-velocity space,
while averaging over the same \ang{12} longitude range as used in our
data analysis. This is, expectedly, similar to the ridge seen in CO,
which is shown in Fig.~\ref{fig:lvd} as colour scale overlay for
comparison \citep[from][]{dame01:_milky_way_co}. Clearly, in
high-spatial-resolution CO data, additional Galactic features can be
resolved, such as the peculiar motions in the nuclear disk close to
the centre of the Galaxy. The kinematics of molecular gas displays
dominant features along the Galactic ridge, beyond this peculiar
motion in the central $\sim$~few \SI{100}{\pc} at rather high
velocities.  But our \Al results do not follow these expectations
along the Galactic ridge, hence clearly the hot, ejecta-carrying gas
does not move as molecular gas does, on these larger
scales. Apparently, \Al carrying interstellar gas moves at
systematically-higher velocities on a large scale than does the
CO-traced molecular gas along the ridge of the Galaxy. For longitudes
$|l| \ga \ang{20}$, its average velocity even exceeds the terminal
velocity of CO, the highest radial velocity seen at a given longitude
\citep{englmaier99:_gas_milky_way}.

What about the influence of the inner bar in our Galaxy?  The dotted
red line shows what would be expected if \num{1/3} of the \Al was
distributed along the Galaxy's long bar, and \num{2/3} of \Al
distributed throughout the disk and spiral arms as above. Here, the
observed slope of the longitude-velocity signature of \Al is
reproduced in the inner part, but apparently \Al kinematics still is
characteristically different outside the regions of the Galaxy's bar
itself, and inner spiral arm regions are involved.

The dashed green line in Fig.~\ref{fig:lvd} combines spiral-arm
sources outside a radius $r_0$ at large-scale galactic rotation with a
new leading-edge blow-out of \SI{\sim 200}{\km\per\s}, which we
suggest is an essential part of explaining the \Al kinematics. We
describe this model and variants in the following subsections.
Apparently, a bar-like distribution of sources could reproduce our
data only for the inner longitude range, while a model based on two
spiral arms extending from the tips of the bar, with large-scale
rotation and a leading-edge blow-out at \SI{200}{\km\per\s}, provides
a closer match to the data and explains the general longitude-velocity
trace as observed in \Al gamma-rays.

In the following subsections, we therefore provide more detail on such
model variants.

%
\subsubsection{Two-arm spiral models}

A simple first-order model to better explain the kinematic properties
of the observed \Al emission is based on the following assumptions:
The spatial distribution of \Al in the Galaxy is along a two-arm
spiral structure as derived from density wave theory (see below for a
four-arm model):
\begin{equation}
  \phi = \frac{1}{\tan(i)} \log \left( \frac{r}{r_0}
  \right) \ \text{for galactocentric radii}\ r > r_0
  \label{eq:spiral}
\end{equation}
where $\phi$ and $i$ denote the azimuth and pitch angle, respectively,
and galactocentric radius $r_0$ defines the inner end of the spiral
arms, which also are assumed to constitute the outer ends of the
Galactic bar. The bar itself does not contribute to the emission in
this model, it only defines the points where the spiral arms begin. We
assume $R_0 = \SI{8}{\kpc}$ for the distance of the Sun from the
Galactic centre. From observations of stars and gas, the spiral-arm
pitch angle has been constrained \citep{francis12:_bisymm_milky_way}
to \ang{5.56}. The \Al emission is assumed to originate from a
\SI{0.5}{\kpc} thick layer around the above-defined spiral arms as
emission zones (the model results below are not sensitive to
reasonable variations of this number), and declines from inner to
outer arm regions as a power law in azimuth angle, i.e.\
$\sim\phi^a$. For the intrinsic velocity of \Al nuclei at their decay,
we adopt an additional azimuthal motion (the blow-out velocity
$v_\mathrm{bo}$), on top of the rotational velocity, which is adopted
as \SI{250}{\km\per\s} everywhere \citep[and discussion in
\citealp{dobbs12:_myth_mol_ring}]{reid09:_trigon_paral_mass_sfreg}.
Additionally we parametrize the bar angle $\alpha$, i.e.\ the angle
between the line from the Galactic centre to the Sun and the line from
the Galactic centre to the near end of the bar, as seen from the
Galactic centre: it is taken to be \ang{38}, towards the upper end of
the range reported by different studies
\citep{francis12:_bisymm_milky_way,
  green11:_major_struc_inner_galax_delin, long13:_made_iii,
  wang12:_milky_way, martinez-valpuesta11:_unify_boxy_bulge}. Smaller
bar angles tend to decrease the fit quality and increase the blow out
velocity by about \SI{10}{\percent}. Our assumption on the bar angle
is consistent with recent measurements from star counts:
\citet{wegg13:_bulge_red_clump} find an angle of \ang{27+-2}. There is
the possibility that the outer bar is twisted \citep[the so-called
long bar, see][]{benjamin05:_first_glimpse,
  cabrera-lavers08:_ukidss_survey}, e.g.\ the bar could have leading
ends \citep{martinez-valpuesta11:_unify_boxy_bulge}. The spiral arm
ends could be leading the bar even more. In fact, our bar angle is
defined by the spiral arm ends and not major axis of the bar.  Our
model is vertically unresolved as our data are (latitude bin width
\ang{10}); for comparison, the nearest spiral arm at \SI{\sim 3}{\kpc}
distance extends over \num{\sim 1/2} of our latitude bin width for an
assumed \Al scale height of \SI{130}{\pc}
\citep{wang09:_spect_galac_al}.

When we perform a $\chi^2$ minimization of this model for our \Al
longitude-velocity data with free parameters $r_0$, $i$,
$v_\mathrm{bo}$ and $\alpha$, we obtain results as shown in
Table~\ref{tab:spiral-fit}. This includes fitting or fits adopting $i$
as constant $i = \ang{5.56}$, and two different bar angles $\alpha$.

\begin{table*}
  \caption{Best-fit parameters for our two-armed spiral emission model
    with different assumptions about the input parameters. For the
    main paper, we have adopted the model with fixed $\alpha =
    \ang{38}$ and $i = \ang{5.56}$. The rightmost column shows reduced
    $\chi^2$ as a measure of fit quality. We also include our best-fit
    four-arm spiral in the bottom row.}
\label{tab:spiral-fit}
\centering
\begin{tabular}{s s S c S S S}
  \hline\hline
  \multicolumn{2}{c}{Input Parameters} &
  \multicolumn{4}{c}{Fitted Parameters} \\
  \cmidrule(r){1-2}\cmidrule(lr){3-6}
  {Bar angle} & {Pitch angle} & {Bar radius} & {Pitch angle} & {\Al
    drop-off} & {\Al vel.} & {$\chi^2$} \\
  {$\alpha$} & {$i$} & {$r_0\ [\si{\kpc}]$} & {$i$} & {$a$} &
  {$v_\mathrm{bo}\ [\si{\km\per\s}]$} \\
  \hline
   \ang{38} &            & 4.4 & \ang{7.7} & 0.35 & 220 \pm 60 & 1.63 \\
   \ang{38} & \ang{5.56} & 4.6 &           & 0.40 & 225 \pm 50 & 1.79 \\
   \ang{20} & \ang{5.56} & 4.9 &           & 0.65 & 240 \pm 60 & 2.00 \\
   \ang{38} &            & 4.1 & \ang{-0.19} / \ang{10.25} & 0.01
   & 180 \pm 75 & 1.01 \\
\hline
\end{tabular}
\end{table*}

\begin{figure}
  \centering
  \includegraphics[width=\linewidth]{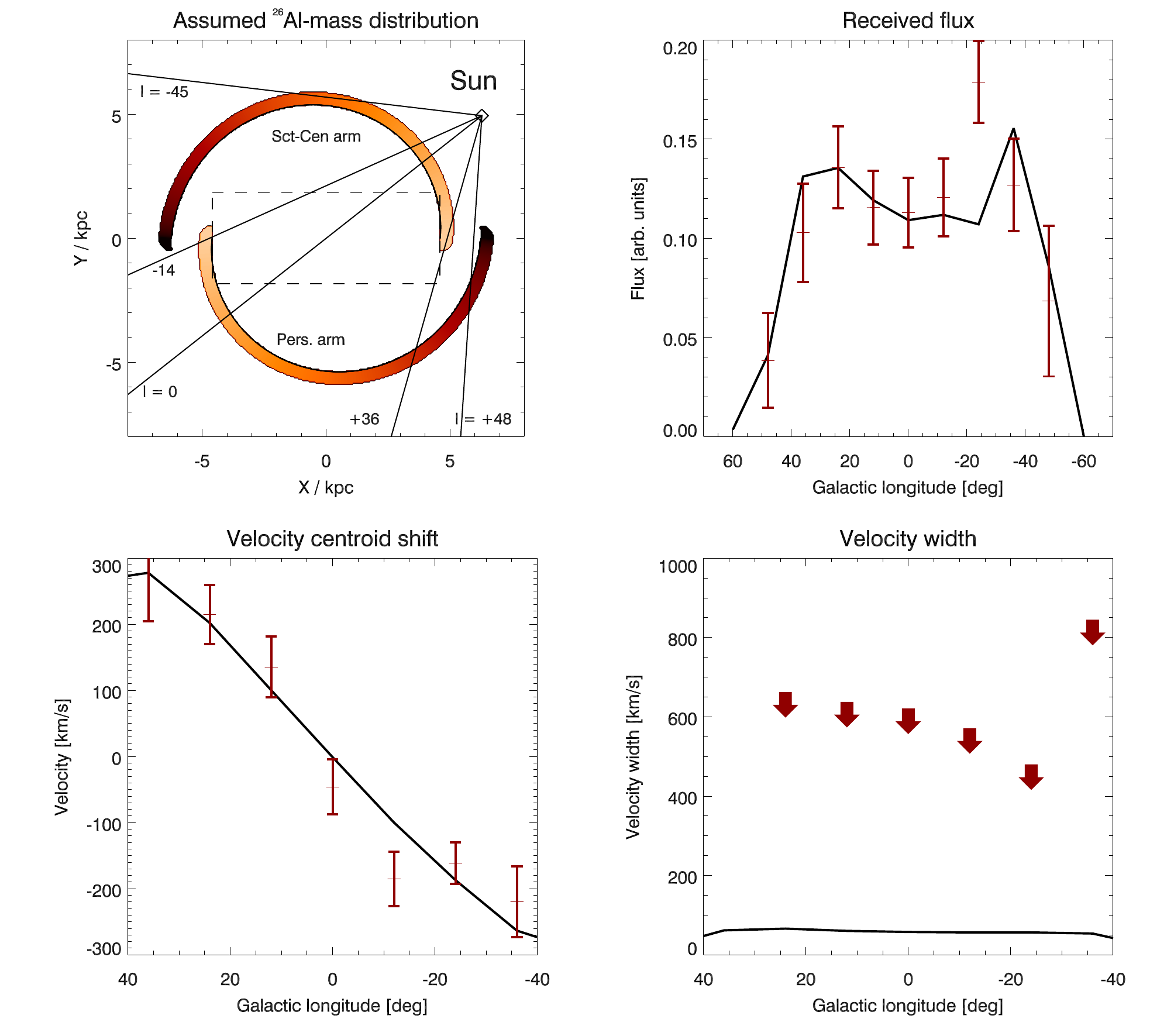}
  \caption{Properties of our preferred model used in the main
    text. The top left plot shows the adopted \Al-source distribution
    in the plane of the Milky Way as seen from above. Also indicated
    is the position of the Sun (\SI{8}{\kpc} from the Galactic
    centre), the location of the bar (dashed) and several longitudinal
    directions of interest. The colour indicates the \Al-density
    (yellow: highest, black: lowest). The other plots show
    quantities derived from this source distribution as a function of
    Galactic longitude as solid black lines (top right: received flux
    in arbitrary units, bottom left: velocity centroid shift, bottom
    right: velocity width). Overplotted are the observed data points
    with 1-standard-deviation-sized error bars, and
    3-standard-deviation upper limits in the case of the velocity
    width. The flux excess at $l = \ang{-24}$ may be due to foreground
    emission from the Sco-Cen association
    \citep{diehl10:_radioac_al_scorp_centaur} leaking from higher
    latitudes into our ROI bin. The observational limits on the
    velocity width are not constraining and compatible with
    expectations (see Sect.~\ref{sec:spi-gamma}).}
  \label{fig:prop-2spiral}
\end{figure}

\begin{figure}
  \centering
  \includegraphics[width=\linewidth]{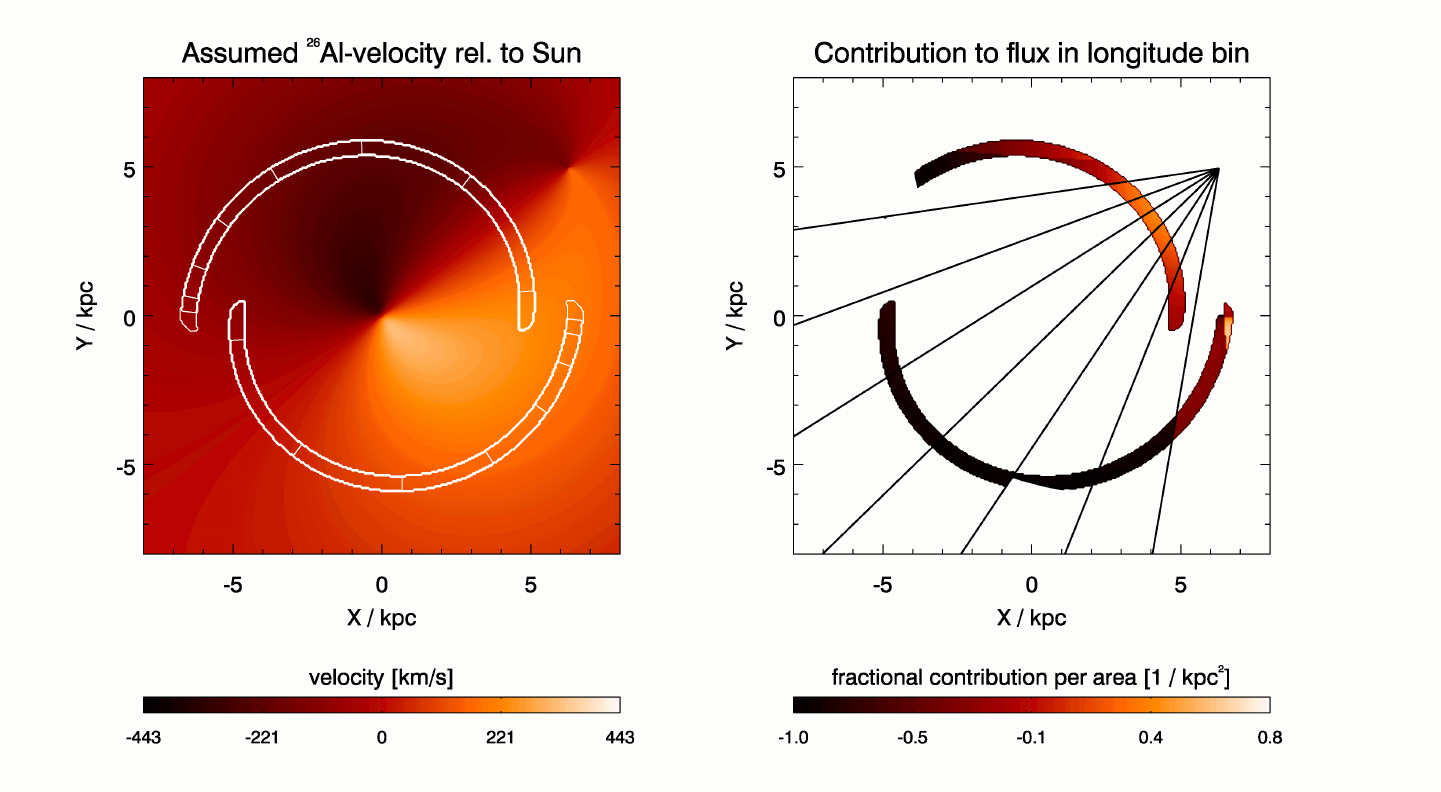}
  \caption{Left: Map of the local velocity relative to the Sun in our
    preferred model. The white contours indicate the \Al mass
    distribution. Right: Relative flux contribution per surface area
    in the standard model, separate for each longitude bin. Several
    longitude bins are indicated by black lines.}
  \label{fig:vrad-2spiral}
\end{figure}

We show the relative contributions along each line of sight of the
assumed source distribution for the preferred two-arm model
(Fig.~\ref{fig:prop-2spiral}) in
Fig.~\ref{fig:vrad-2spiral}. Effectively, the foreground part of the
source emission along each line of sight dominates the observed
signal. Towards the far bar end, the velocity of the \Al carrying gas
turns out to be similar to the velocity of gas in the foreground
arm. Moreover, the foreground arm dominates in brightness. Hence, the
far end of the bar cannot be disentangled.

In summary, the blowout velocity depends only very weakly on changes
to any of the parameters of the model. For the results of our paper we
have adopted the fit for a fixed $\alpha = \ang{38}$ and $i =
\ang{5.56}$, and blowout velocity $v_\mathrm{bo} = \SI{225 \pm
  50}{\km\per\s}$, because $\chi^2$ is only marginally worse, and for
consistency with a recent study
\citep{francis12:_bisymm_milky_way}. This model is shown in
Fig.~\ref{fig:prop-2spiral}.

%
\subsubsection{Four-arm spiral models}

\begin{figure}
  \centering
  \includegraphics[width=\linewidth]{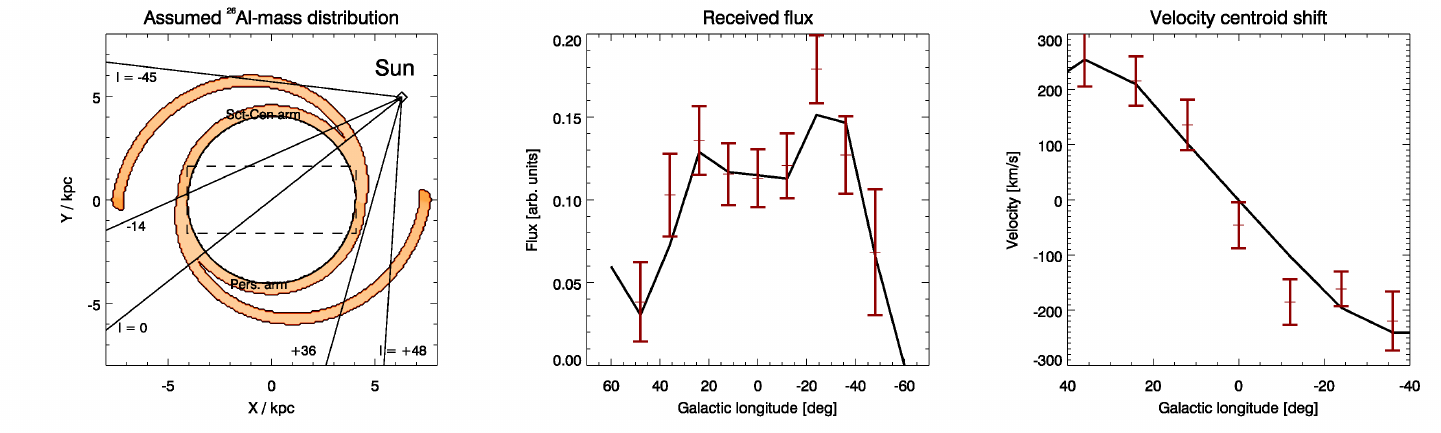}
  \caption{Similar to Fig.~\ref{fig:prop-2spiral}, but for the best-fit four-arm
    spiral.}
  \label{fig:prop-4spiral}
\end{figure}

We also investigated four-arm spiral models for the Galaxy, with
characteristics similar to the two-arm model.  The best-fit four-arm
model (Fig.~\ref{fig:prop-4spiral}) leads to a reduced $\chi^2$ of
\num{0.93}, thus formally fits much better than the two-arm model.
However, the improved fit quality is mainly due to a better, but still
not fully satisfactory match of the flux at $l = \ang{-24}$, which we
believe to be related to foreground emission. We therefore prefer the
best-fit two-arm model over the four-arm one, because it has fewer
parameters. The inferred blow-out velocity depends weakly on this
choice (compare Table~\ref{tab:spiral-fit}).

%
\subsection{Comparison to other studies}

%
\subsubsection{Spiral arm structure}

\begin{figure}
  \centering
  \includegraphics[width=\linewidth]{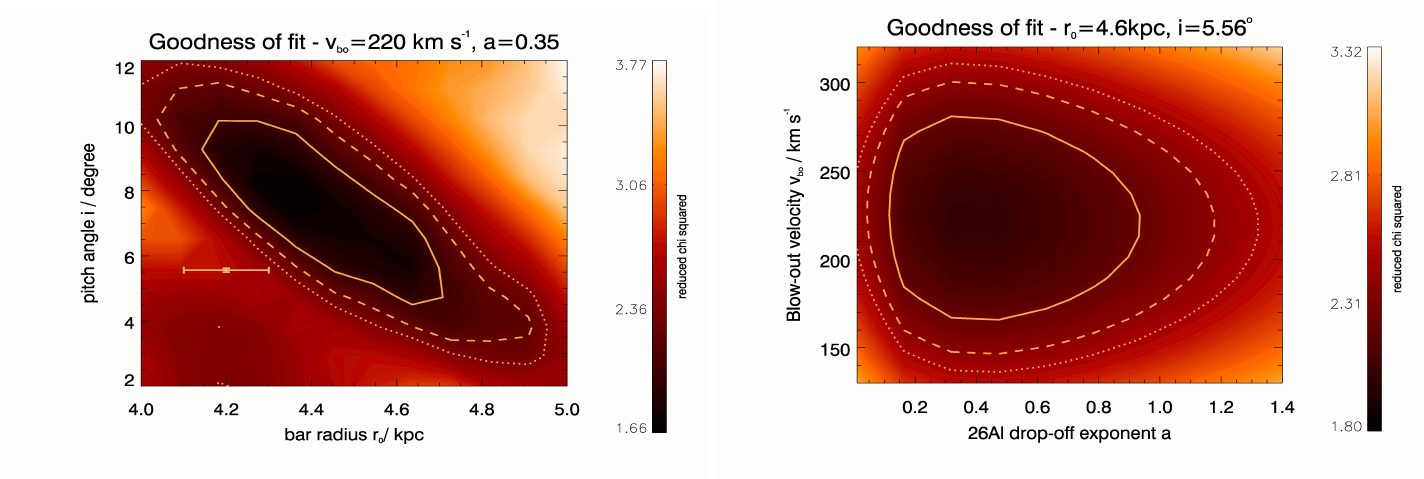}
  \caption{Reduced $\chi^2$ and confidence contours as a function of
    galactocentric bar radius and pitch angle for the Galactic spiral
    structure (left, the value from
    \citet{francis12:_bisymm_milky_way}. is also indicated, including
    1-standard-deviation error bars) and as a function of the
    \Al-drop-off exponent and the blowout velocity (right). The
    \SIlist{68;90;95}{\percent} confidence levels are shown by the
    solid, dashed and dotted lines. For the left plot, the overall
    best-fit values of $a = \num{0.35}$ and $v_\mathrm{bo} =
    \SI{220}{\km\per\s}$ have been kept fixed. For the right plot, the
    pitch angle from \citep{francis12:_bisymm_milky_way} $(i =
    \ang{5.56})$ and the best-fit bar radius for fixed $i =
    \ang{5.56}$, $r_0 = \SI{4.6}{\kpc}$, has been used (see also
    \citet{martinez-valpuesta11:_unify_boxy_bulge}). The blowout
    velocity depends only weakly on these parameters (compare
    Table~\ref{tab:spiral-fit}).}
  \label{fig:conf-cont}
\end{figure}

Confidence levels for two respective pairs of our model parameters are
shown in Fig.~\ref{fig:conf-cont}. Clearly, a pitch angle of
\ang{5.56} is consistent with our data, within uncertainties. Our
favoured Galactic-bar radius falls into the range of published values
\citep{francis12:_bisymm_milky_way,
  martinez-valpuesta11:_unify_boxy_bulge}.

%
\subsubsection{Gas components in our and other galaxies}

The blow-out velocity that we derive is inconsistent with velocity
variations expected at spiral shocks in the context of simple density
wave theory without explicit consideration of the interaction of
massive star feedback with the underlying density structure: Our
obtained velocities significantly exceed the ones found in molecular
\citep{dame01:_milky_way_co} and atomic gas
\citep{kalberla08:_global_h_i_milky_way}. Gas-kinematics models
\citep{fux99:_n_milky_way_gas, bissantz03:_gas_milky_way,
  baba10:_inter_lv_milky_way, khoperskov13:_milky_way} are in good
agreement with these observations, but do also not produce velocities
which come close to the ones we
find. \citet{shetty07:_kinem_spiral_arm_stream_m51} extract the
azimuthal velocity variations from CO data for the galaxy M51, which
features very prominent spiral arms. Even for the stronger spiral arm
in M51, the azimuthal velocity shear does not exceed \SI{\approx
  120}{\km\per\s}. The massive stars are however observed downstream
of the maximum of the CO intensity, wherefore only a small fraction of
the velocity shear would appear as \Al-velocities in M51. Further,
most of the velocity shear in M51 is due to the slowing down of the
gas at the spiral shock, such that the azimuthal velocity drops
significantly below the circular velocity. Our observations require
however velocities of at least \SI{130}{\km\per\s} (three standard
deviations) above the circular velocity. The picture derived by
detailed molecular gas observations for M51
\citep{egusa11:_molec_gas_evolut_acros_spiral_arm_m51} would explain
our data well: Molecular clouds enter the spiral shock from behind and
merge there to larger clouds. This triggers the formation of massive
stars downstream of the density maximum. Thus the massive star ejecta
are impeded by the upstream density maximum and obtain a net velocity
into the direction of rotation.

Our results imply that blow-out occurs from the star-forming regions
of the spiral arms within the plane and plausibly also towards the
halo. We expect star formation to occur as gas falls into the
spiral-arm potential
\citep{wada11:_galac,athanassoula92:_dust_lanes}. Observed young star
clusters tend to be associated with spiral arms, particularly towards
their inner ends and at the near
\citep{lopez-corredoira99:_major_star_form} and far
\citep{davies12:_galac_bar} ends of the Galaxy's bar. Simulations
\citep{athanassoula12:_towar_galax} and face-on images
\citep{elmegreen12:_trigger} of barred spiral galaxies confirm this
general picture. Inside the corotation radius, objects move faster
than the large-scale patterns of the bar and of the spiral
arms. Consequently, gas clouds which enter the pattern and form stars
will create stellar groups that appear offset towards the leading edge
of the pattern. For the Galactic bar, a corotation radius of \SI{\sim
  4}{\kpc} has been found \citep{gerhard11:_patter_milky_way}, though
estimates increase to \SIrange[range-phrase = { or even }]{5}{6}{\kpc}
if both a boxy bulge and bar are considered
\citep{martinez-valpuesta11:_unify_boxy_bulge}. A corotation of
\SI{\sim 8.4}{\kpc} has been found for the spiral arms of our Galaxy
\citep{lepine11:_galax_cs}. For a pattern velocity difference of
\SI{\sim 50}{\km\per\s}, an offset of \SI{\sim 200}{\pc} between
superbubble-blowing massive stars and spiral-arm density maximum would
be expected at the end of a typical massive-star lifetime of
$\sim$~few \SI{e6}{years}. The \Al sources typically located in a
density gradient on one side of the large-scale density enhancement in
arms would trace non-isotropic superbubble growth around massive star
groups, which otherwise cannot easily be seen in remains from parental
molecular clouds
\citep{louie13:_geomet_offset_spiral_arms_m51}. Although systematic
velocity variation across spiral arms is expected from density wave
theory and has been found in M51
\citep{shetty07:_kinem_spiral_arm_stream_m51}, such velocity
differences across a spiral arm are only \SI{\sim 100}{\km\per\s} or
lower, and cannot explain our measurement (see
Sect.~\ref{sec:sk-model}).  Simulations show that the initially
isotropic ejecta from massive stars face and enhance the high-pressure
region on one side, and a champagne-like outflow into the opposite
direction occurs within a rather short time
\citep{fierlinger12:_molec_cloud_disrup_chemic_enric,
  baumgartner09:_metal}. Our observations require that \Al-enriched
superbubbles preferentially expand in the direction of galactic
rotation relatively unimpeded, whereas they are blocked by denser gas
in the opposite direction. This probably occurs along spiral arms and
near the tip of the bar where the two prominent inner spiral arms
curve towards the bar.  Interestingly, velocities around
\SI{200}{\km\per\s} are close to the sound speed in the hot gas
(\SI{e6}{\K}). \Citet{glasow13:_galac_lyman} model the absorption
systems seen at similar velocities against Lyman break galaxies as
cooled-down shells from expanding superbubbles. In both cases, the
velocity would be set by the sound speed of the hot phase the
superbubbles are expanding into.

%
\subsubsection{Hot gas in and above the Galactic plane}

Similar velocities perpendicular to the Galactic plane are also
consistent with the scale height of \Al measured
\citep{wang09:_spect_galac_al} as $130^{+120}_{-70}$~\si{\pc}: The
characteristic scale height of parental molecular clouds is about
\SI{50}{\pc} \citep{dame01:_milky_way_co}, while, for example, a
velocity of \SI{200}{\km\per\s} implies that a height of about
\SI{200}{\pc} above the parental clouds is reached within the decay
time of \SI{e6}{years}. Such velocities cannot support a wind from the
Galaxy into the intergalactic medium. Since \Al traces superbubbles
with ongoing input by massive stars, these should exhibit the highest
velocities, whereas older bubbles that are not highlighted by \Al will
be less dynamic. Hot gas at higher latitudes above the Galactic disk
is also seen in \ion{O}{vi} absorption
\citep{sembach03:_highl_ioniz}. Such high ionization states appear in
hot and relatively dense gas, as the superbubble wall interacts with
ambient gas. \ion{O}{vi} data reflect more closely the average hot gas
in general, including the later stages of superbubble evolution: the
observed \ion{O}{vi} height is \SIrange{2}{3}{\kilo \parsec}
\citep{sembach03:_highl_ioniz}.

Within the Galactic disk, the velocities inferred from absorption also
do not directly follow the molecular gas velocities, but are closer
\citep{sembach03:_highl_ioniz} to them than to the higher velocities
we infer for \Al. Therefore, asymmetries that we observe during the
active superbubble phase apparently are dissipated, and the flow is
more isotropic at later times.

%
\section{Conclusions}
\label{sect-conclusions}

Our measurements of the Doppler shifts of the \SI{1808.63}{\keV} line
as a function of Galactic longitude show that the radial velocity of
the interstellar gas containing \Al in the inner Galaxy differs
significantly from that of other components of the ISM such as those
seen in CO or \ion{H}{I}. We observe the same qualitative behaviour:
there is almost no average radial motion in the direction toward the
the Galactic centre, the magnitude of the radial velocity increases
with the angular separation from the centre up to $|l| \sim \ang{30}$
and the sign of the radial velocity is positive (redshift) in the
first and negative (blueshift) in the fourth Galactic
quadrant. However, the absolute velocities of the \Al-carrying gas are
much larger. Since the line emission happens over a large radial
velocity range, the total emission from the inner Galaxy is
Doppler-broadened. Our measurements of the line width of the \Al
emission are compatible with this implication. The variation of our
measured radial velocity values with the extent of the
region-of-interest we average over is comparable to the statistical
uncertainties and the systematic uncertainties related to the time
dependent variation of the instrument's spectral response are small in
comparison. Since the majority of \Al is produced in massive stars, we
conclude that our observations are probably due to large-scale
asymmetric outflows from the regions where massive stars have formed
recently.

The inner spiral arms, which are the plausible source regions
producing most of the \Al, show blow-out of their massive-star ejecta
preferentially into the direction of the leading edges. This increases
\Al velocities to \SI{\sim 200}{\km\per\s} in addition to large-scale
galactic rotation. Superbubbles are expected to form around massive
star groups on the \si{\mega\year} time scale
\citep{krause13:_feedb}. In a wind-blown cavity or supernova remnant,
initially freely-travelling ejecta would be decelerated within a time
much shorter than a radioactive-decay lifetime. The \Al-rich ejecta
accumulate behind the swept up ambient medium, expanding at a similar
velocity. When estimating the line-of-sight averaged velocities of \Al
as it decays, we should distinguish between the velocity of the bubble
expansion itself and the velocities of ejecta flows within a bubble.
Bubbles from single stars are expected to reach sizes on the order of
\SI{10}{\parsec} with associated expansion velocities of order
\SI{10}{\km\per\s}. Our inferred expansion velocity of \SI{\sim
  200}{\km\per\s} shows that the bubbles highlighted by \Al have to be
powered by many, hence clustered, massive stars. While we defer
detailed consistency checks via cluster-population synthesis to future
work, our measurement together with the fact that \Al ejection is
strongly correlated to the energy injection \citep{voss09:_using_ob}
suggests that massive star feedback in the inner few \si{\kpc} of the
Galaxy and its spiral arms is dominated by sizeable star clusters
producing superbubbles, constituting a fundamental unit of large-scale
stellar feedback.  This does not imply expansion of the superbubble as
an entity with such velocities, but rather provides a measurement from
its interior reflecting its size and position with respect to the \Al
sources.

Our measurements reveal new aspects of large-scale gas kinematics in
the Galaxy, derived from data originating in the hot and tenuous phase
of the ISM that is otherwise hard to measure.  Flow asymmetries
require distinct structure within the spiral arms of the Galaxy: dense
gas, which marks the spiral-arm potential, must be offset upstream
from massive stars, and these must be located towards the spiral-arm
leading edges, with ejecta blowing out into the inter-arm low density
environment (and halo). Origins of spiral arms are debated
\citep{wada11:_galac}, and massive star offsets from their gas density
maxima remain controversial
\citep{louie13:_geomet_offset_spiral_arms_m51,
  ferreras12:_swift_uvot_ngc}. The excess velocity of \Al-traced gas
over stars and cold/dense gas (Fig.~\ref{fig:lvd}) thus constitutes a
clear, though indirect, demonstration of the offsets between
star-forming gas and young stars, which seemed plausible in
density-wave theory and have also been found in molecular gas
observations \citep{vogel88:_star} and images of galaxies
\citep{elmegreen12:_trigger} (see also Sect.~\ref{sec:sk-model}).

This one sided blow-out will impart a local braking torque on the cold
gas as it rotates in the plane of the Galaxy. Once injected into the
halo, hot gas will likely exchange its angular momentum with the
\SI{e6}{\K} corona \citep{gupta12:_huge_reser_ioniz_gas_milky_way}
over \SI{\sim e8}{\year} before cooling and returning to the Galactic
disk elsewhere \citep{marinacci11:_galac_fount}. The total mass of \Al
in the Milky Way is measured \citep{diehl06:_radioac_al_galax} to be
\SI{\sim 2}{\msun}.  This traces about \SI{e6}{\msun} of total
massive-star ejecta, hence a hot gas flow of \SI{\sim 1}{\msun \per
  \year}, ejected into the direction of Galactic rotation, with an
excess velocity comparable to the Galactic rotation velocity
itself. These \Al data measure primarily a torque of a specific, and
otherwise hard to observe, gas component, which also couples to the
total of Galactic gas flows. The global recoil may slow down denser
gas in its rotation in the Galaxy: Its global torque of $\SI{1}{\msun
  \per \year} \cdot \SI{4}{\kpc} \cdot \SI{200}{\km\per\s} \approx
\SI{e6}{\msun \kpc \squared \per \mega \year \squared}$ can be
compared to the total angular momentum of the Milky Way's gas of
roughly $\SI{e10}{\msun} \cdot \SI{5}{\kpc} \cdot \SI{250}{\km\per\s}
\approx \SI{e10}{\msun \kpc \squared \per \mega \year}$. To estimate
potential impacts, this blow-out could remove the entire angular
momentum from the dense gas within \SI{10}{\giga\year}. This estimate
assumes that the angular momentum would not return to the disk. More
realistically, some angular momentum exchange with the gaseous halo
will take place, and some fraction of the angular momentum will return
to the disk when the ejecta fall back. Estimated radial inflow rates
toward the inner Galaxy for other processes are
\citep{crocker12:_nontherm_galac_centr_fermi} \SIrange{0.1}{1}{\msun
  \per \year} due to gravitational or \SI{0.2}{\msun \per \year} due
to magnetic torque, and \SI{0.2}{\msun \per \year} by mass loss from
the bulge stars.  Radial gas diffusion is implied by one-sided
superbubble blowout even if there was no global loss of angular
momentum from the Galactic disk, because it would require substantial
fine-tuning of the angular momentum-exchange between the off-streaming
\Al-traced gas and the halo gas to get the \Al-traced gas back to its
original position.  One-sided superbubble blow-out may thus contribute
both to linking general star formation on \si{\kpc} scales to
large-scale gas flows and to subsequent star formation in the inner
regions of our Galaxy.  Our measured characteristic ejecta velocity
suggest that superbubbles are the dominant structure of the ISM around
massive stars.

The UV luminosity associated with massive stars is important in
ionizing gas at higher Galactic latitudes and towards the halo
\citep{razoumov06:_escap_ioniz_radiat}: It's propagation and escape
from the denser parts of the disk depends critically on the structure
of the interstellar medium around its sources and the erosion of
spiral-arm gas around massive stars. The morphology of surrounding
interstellar gas determines the extent to which UV from massive stars
may reach more distant gas, and contributes to the ionization state of
the intergalactic medium. Blow-out from massive star regions in
superbubbles is thus a fundamental aspect of large-scale stellar
feedback in a star-forming galaxy such as our own. This effect must be
accounted for in models \citep{veilleux05:_galac_winds} of galactic
winds and galaxy evolution, which currently consider either single,
non-interacting bubbles, or the galaxy as a whole with simplified
structural assumptions at scales below $\sim$~\si{\kpc} scales.

%
\begin{acknowledgements}
  This research was supported by the German DFG cluster of excellence
  ``Origin and Structure of the Universe''. The INTEGRAL/SPI project
  has been completed under the responsibility and leadership of CNES;
  we are grateful to ASI, CEA, CNES, DLR, ESA, INTA, NASA and OSTC for
  support of this ESA space science mission.
\end{acknowledgements}

%
\bibliographystyle{aa}
\bibliography{al26-galaxy-asym}

%
\appendix

%
\section{Systematic uncertainties}
\label{sec:lvd-analysis_errors}

\begin{figure}
  \centering
  \includegraphics[width=\linewidth]{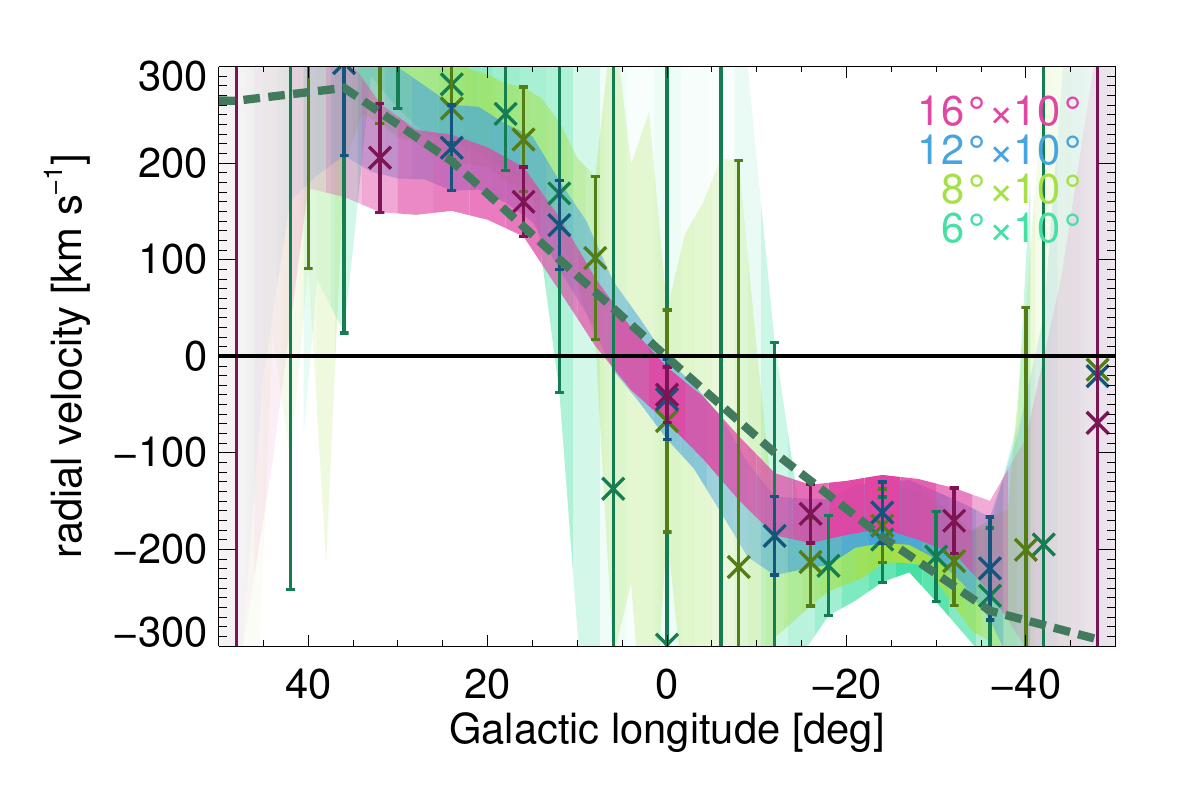}
  \includegraphics[width=\linewidth]{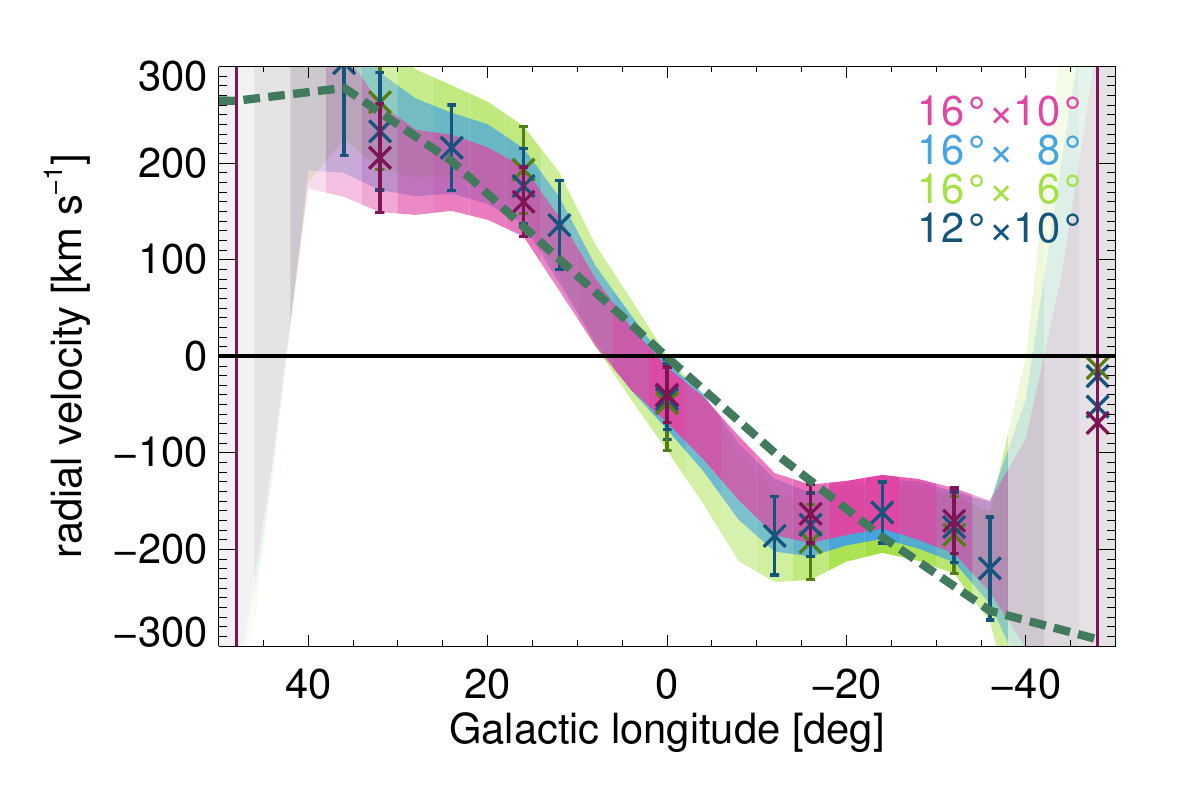}
  \caption{\Al longitude-velocity results for different bin sizes
    in Galactic longitude {\it (above)} and latitude {\it (below)}.}
  \label{fig:lvd-dl}
\end{figure}

We investigated the impact of choosing sky region bins of different
sizes, e.g.\ smaller at the expense of a smaller signal, but aiming to
obtain a better spatial resolution for our velocity information
(spectra in Fig.~\ref{fig:spimodfit-sa-4spec} illustrate this
specifically towards longitude $l \sim {}$\SIrange{-27}{-28}{\degree},
the brightest part of the 4\textsuperscript{th} Galactic quadrant).
Figure~\ref{fig:lvd-dl} shows the dependence of the \Al velocity
results on the ROI extent in longitude (varying between \ang{6} and
\ang{16}), and in latitude (between \ang{6} and \ang{10}).  We can
obtain higher spatial resolution, but only in regions of high
intensity.  In our longitude-velocity figures, we include the Bayes
factor information for $B > 1$ also for the data points where
detections are insignificant, and the data point values therefore are
not shown themselves in the figure.  In order to maximize the \Al
signal for best determination of its line centroid position, we chose
a latitude range of \ang{+-5} for our definite analysis.

\begin{figure}
  \centering
  \includegraphics[width=\linewidth]{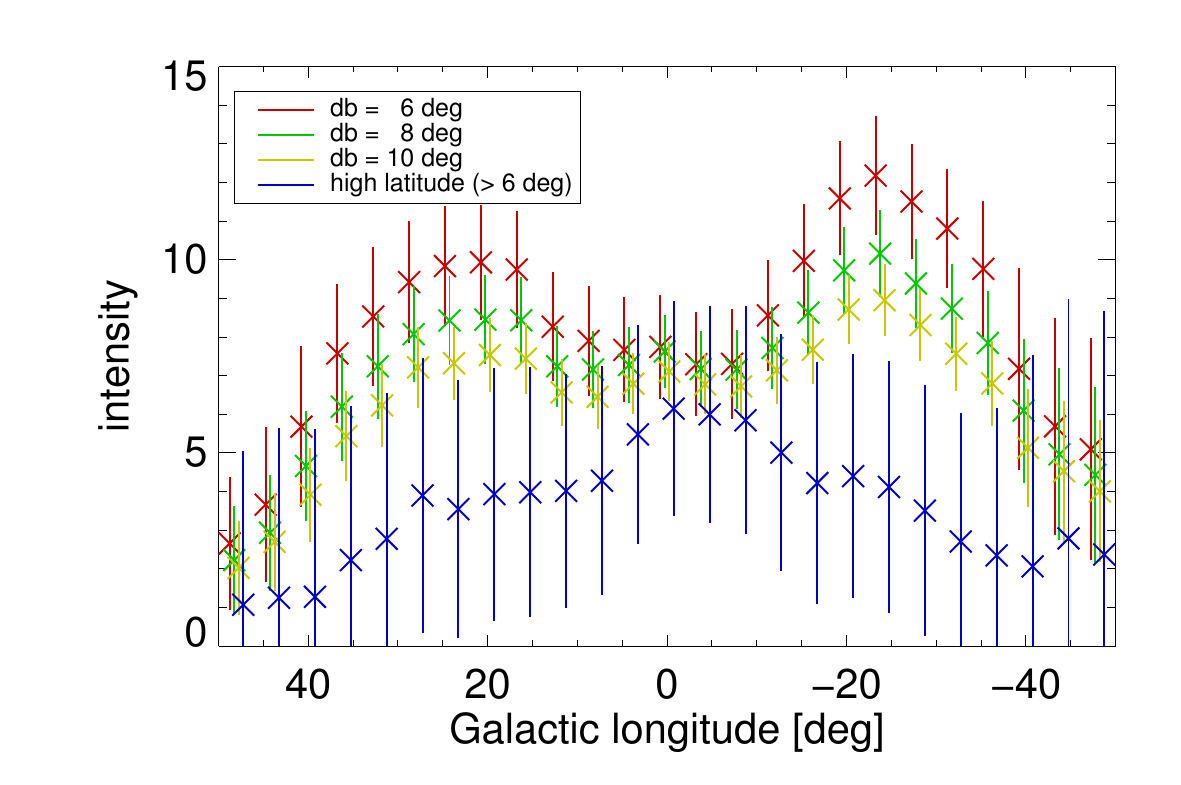}
  \caption{\Al intensities along the inner Galaxy, for different
    bin sizes in Galactic latitude, and \ang{16} bins in longitude,
    offset in \ang{4} steps. The emission which can be attributed to
    higher latitudes, hence potential foreground sources, is indicated
    in the lowest (blue) data points, derived from the difference in
    intensities for the \ang{6} and \ang{10} bins.}
  \label{fig:inten-db}
\end{figure}

The variation of intensity (per solid angle, averaged over the ROI)
with latitude extent offers some indication that the intensity falloff
with latitude is slower in the inner ($|l| < \ang{10}$) Galactic
plane. Figure~\ref{fig:inten-db} shows these intensity variations for
latitude extents of \SIlist{6;8;10}{\degree} as well as the difference
between \ang{10} and \ang{6} ROIs as ``high latitude'' (units are
relative).  We can see that the intensity distribution varies
consistently for different latitude integration regions of
\SIrange{6}{10}{\degree}, except for a possible additional component
between longitudes \ang[retain-explicit-plus]{+5} and \ang{-12}. This
component may arise from the Scorpius-Centaurus region
\citep{diehl10:_radioac_al_scorp_centaur}.  This different behaviour
in the inner Galaxy also is seen in the shape of the \Al line: It
appears that we see a superposition in the region $l \sim [\ang{-15},
\ang{10}]$ (see Fig.~\ref{fig:inten-db}) of a nearby (and more
extended in latitude) component which is approximately at rest with
respect to the observer, and the Doppler-shifted component from the
Galactic plane. Therefore, the line shape is less well represented by
a single Gaussian, and the uncertainties in centroid determination are
larger in this inner region of the Galactic plane (see
Fig.~\ref{fig:lvd-dl}). (Additionally, in $l \sim [\ang{-15},
\ang{15}]$ the line is less bright than at $|l| \sim [\ang{15},
\ang{30}]$, further increasing the line centroid's uncertainty).

Detailed tracking of the spectral response variation with degradations
and annealings has been found necessary for determining the width of
the \Al line (as shown in Fig.~\ref{fig:spec-dl128db12}).  But the
effect of the spectral response variation on the line position
measurements for smaller emission regions is small: compare
Fig.~\ref{fig:spec-dl128db12} to Fig.~\ref{fig:spectra-ref}). For
example, with an instrumental line width of \SI{3}{\keV} and a
celestial line width of \SI{1.4}{\keV}
(Fig.~\ref{fig:spec-dl128db12}), the measured width is
$\sqrt{(\SI{3}{\keV})^2 + (\SI{1.4}{\keV})^2} \approx
\SI{3.31}{\keV}$, which is a \SI{10}{\percent} excess above the
instrumental width.

\begin{figure}
  \centering
  \includegraphics[width=\linewidth]{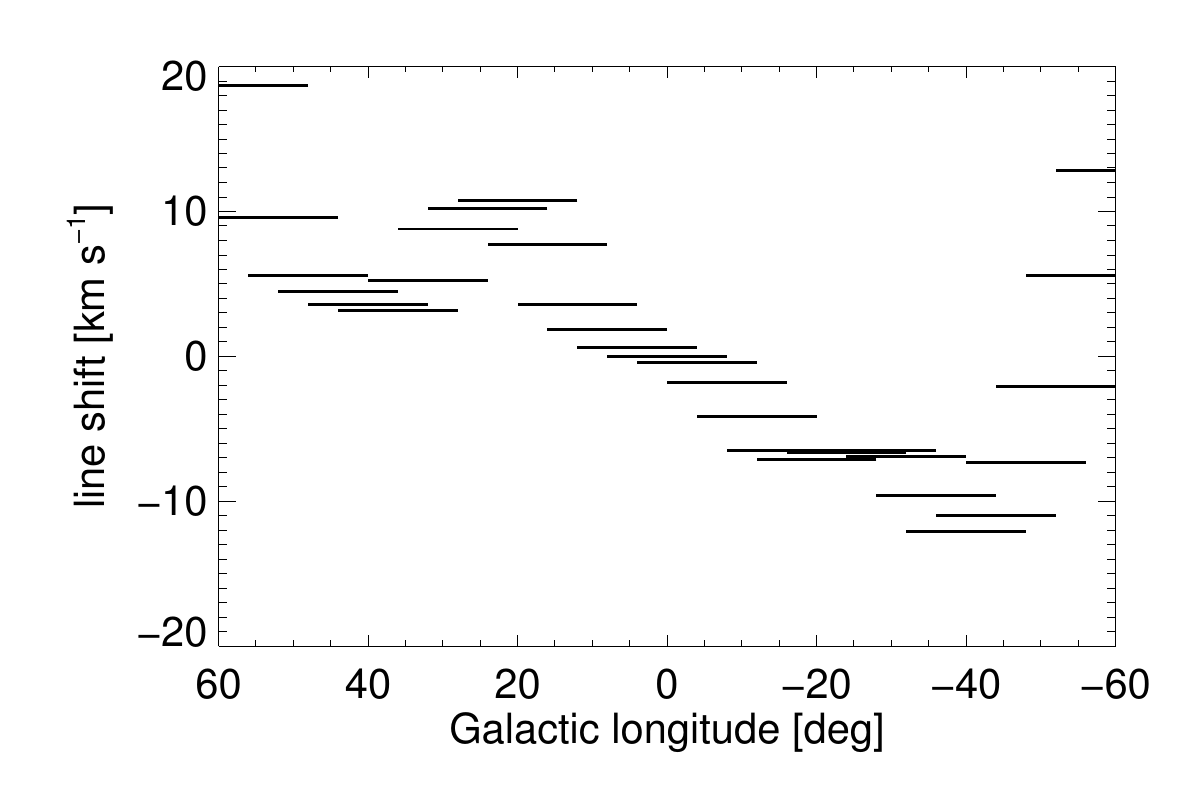}
  \caption{Offset in line centroids from the effect of time-variable
    degradation, if a line with fixed centroid would be represented by
    a Gaussian fit. Periods where the respective longitude range was
    observed are used, so that the impact on our longitude-velocity
    result can directly be seen.}
  \label{fig:resp-avg-long}
\end{figure}

We then investigated this variation for a correlation with the times
of our observations along the plane of the Galaxy, possibly leading to
a longitude-dependent bias. For this, we determine the maximum
position shift in units of velocity, measured by when fitting an
instrumental line to the data that were taken while pointing at the
respective longitude intervals. As shown in
Fig.~\ref{fig:resp-avg-long}, the bias which may result from this
time-variable spectral response is small (\SI{< 10}{\km\per\s})
compared to our reported line-shift values.

\begin{figure}
  \centering
  \includegraphics[width=\linewidth]{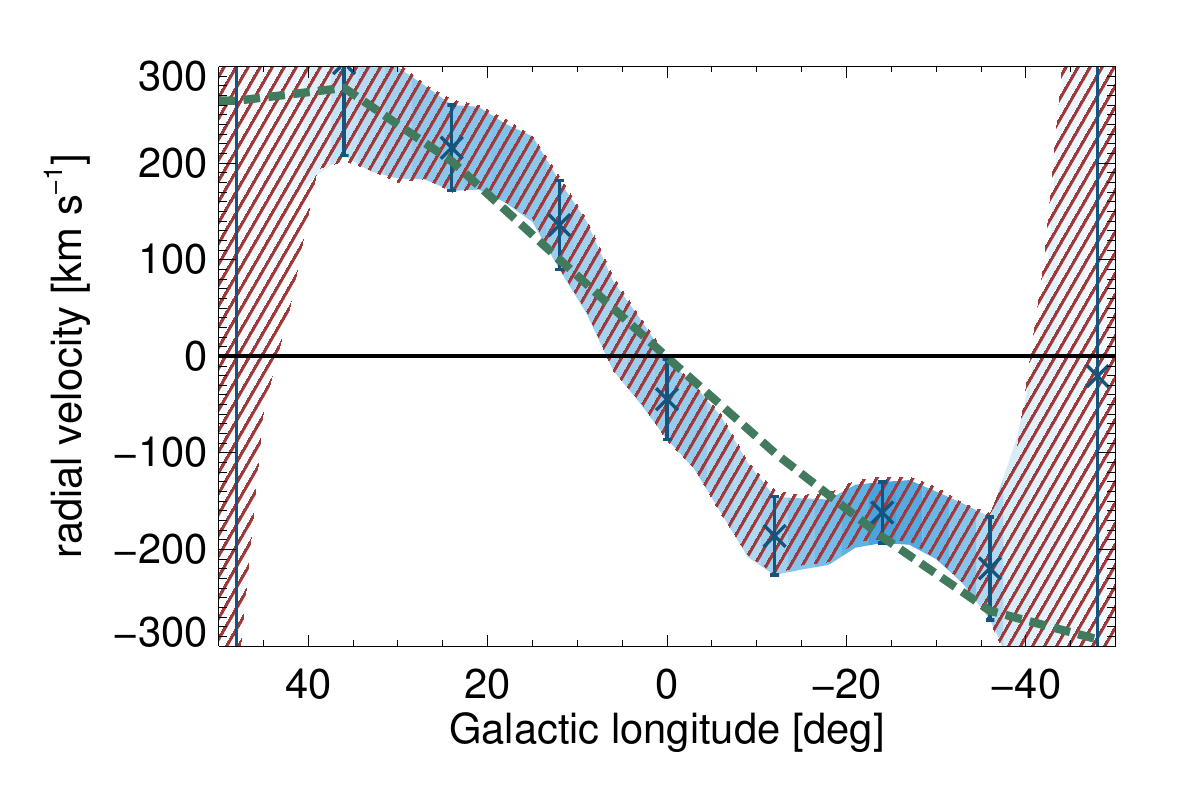}
  \caption{Longitude-velocity results using different methods of
    determining the spectral response (see
    Sect.~\ref{sec:spi-gammashape}).  For comparison we show: Blue
    data points and blue shading in the background: measured
    instrumental background line; red hatched regions in the
    foreground: line shape model fitted to a set of instrumental
    background lines.}
  \label{fig:lvd-resp}
\end{figure}

The observed velocity values for different latitude ranges are all
consistent within the uncertainties.

We conclude that the systematic uncertainties in our
longitude-velocity measurement are smaller than statistical
uncertainties and do not alter the results reported, specifically the
asymmetry of blow-out as discussed. Also our background method and
spectral-response treatment does not have an impact on the
results. Systematics are dominated by the selection of the ROI used,
and may affect the detailed velocity value derived for a particular
region along the Galactic plane by \SI{\sim 50}{\km\per\s}.

\end{document}